\documentclass[twocolumn]{article}

\usepackage{arxiv}

\usepackage{times}
\usepackage{soul}
\usepackage{url}
\usepackage[hidelinks]{hyperref}
\usepackage[utf8]{inputenc}
\usepackage[small]{caption}
\usepackage{graphicx}
\usepackage{amsmath}
\usepackage{booktabs}
\usepackage{algorithm}
\usepackage{algorithmic}
\usepackage{cite}
\usepackage[switch,displaymath,mathlines]{lineno}
\usepackage{xcolor}
\usepackage{comment}
\usepackage{marginnote}
\usepackage{amssymb}
\usepackage{subcaption}

\newcommand{\comb}[2]{{}_{#1}\mathrm{C}_{#2}}
\newcommand{\commentout}[1]{}
\newcommand{\novspace}[1]{}
\newcommand{\conv}{*}

\DeclareMathOperator{\EX}{\mathbb{E}}% expected value

\title{Incorporating Visual Cortical Lateral Connection Properties into CNN: 
\vspace{0.3em}
\\Recurrent Activation and Excitatory-Inhibitory Separation}

\author{
  Jin Hyun Park \\
  Dept. of Computer Science and Engineering \\ 
  Texas A\&M University, College Station \\
  Texas, USA \\
  \texttt{jinhyun.park@tamu.edu} \\
  \And
  Cheng Zhang \\
  Dept. of Computer Science and Engineering \\ 
  Texas A\&M University, College Station \\
  Texas, USA \\
  \texttt{chzhang@tamu.edu} \\
  \And
  Yoonsuck Choe \\
  Dept. of Computer Science and Engineering \\ 
  Texas A\&M University, College Station \\
  Texas, USA \\
  \texttt{choe@tamu.edu} \\
}

\begin{document}

\twocolumn[
    \begin{@twocolumnfalse} % Switch to one-column mode
    
        \maketitle
        
        % Add equal contribution footnote here
        \begin{abstract}
        The original Convolutional Neural Networks (CNNs) and their modern updates such as the ResNet are heavily inspired by the mammalian visual system. These models include afferent connections (retina and LGN to the visual cortex) and long-range projections (connections across different visual cortical areas). However, in the mammalian visual system, there are connections {\em within} each visual cortical area, known as lateral (or horizontal) connections. These would roughly correspond to connections within CNN feature maps, and this important architectural feature is missing in current CNN models. In this paper, we present how such lateral connections can be modeled within the standard CNN framework, and test its benefits and analyze its emergent properties in relation to the biological visual system. We will focus on two main architectural features of lateral connections: (1) recurrent activation and (2) separation of excitatory and inhibitory connections. We show that recurrent CNN using weight sharing is equivalent to lateral connections, and propose a custom loss function to separate excitatory and inhibitory weights. The addition of these two leads to increased classification accuracy, and importantly, the activation properties and connection properties of the resulting model show properties similar to those observed in the biological visual system. We expect our approach to help align CNN closer to its biological counterpart and better understand the principles of visual cortical computation.
        \end{abstract}
    \end{@twocolumnfalse} 
    \vskip 0.4in % Adjust vertical spacing before resuming two-column mode
]

\section{Introduction}

Biologically motivated neural networks for visual processing such as Neocognitron \cite{fukushima1980neocognitron}, Convolutional Neural Networks (CNNs) \cite{lecun1989backpropagation}, and HMAX \cite{riesenhuber1999hierarchical}, drew inspiration from Hubel and Wiesel's works on the primary visual cortical neurons \cite{hubel1959receptive} and subsequent developments in the field. A common feature in these models is that the alternating layers of simple cells and complex cells form a feed-forward hierarchy, starting with the afferent connections from the input (for CNN, the convolutional layers and pooling layers may serve the same purpose \cite{lindsay2021convolutional}). 

The hierarchy in these models loosely mimic the projections among different cortical areas in the visual pathway \cite{felleman1991distributed}, where each convolutional layer correspond to a distinct visual cortical area, and the connections serving as the long-range projections. Functionally, feature representations in CNN also seem to show close similarity to those in the ventral visual pathway \cite{zeiler2014visualizing}. 

There is a major shortcoming in this, since the various visual cortical areas do not form a strict hierarchy, as there are feedback connections between the visual areas forming a recurrent loop \cite{briggs2020role}. Some architectural features in modern CNN variants may serve this purpose. For instance, Liao and Poggio \cite{liao2016bridging} proposed that skipped connections in the ResNet \cite{he2016deep} can be seen as implementing such recurrent projections (also see Recurrent CNN: \cite{liang2015recurrent}).

\begin{figure}[t]
\centering
{\includegraphics[width=0.35\textwidth]{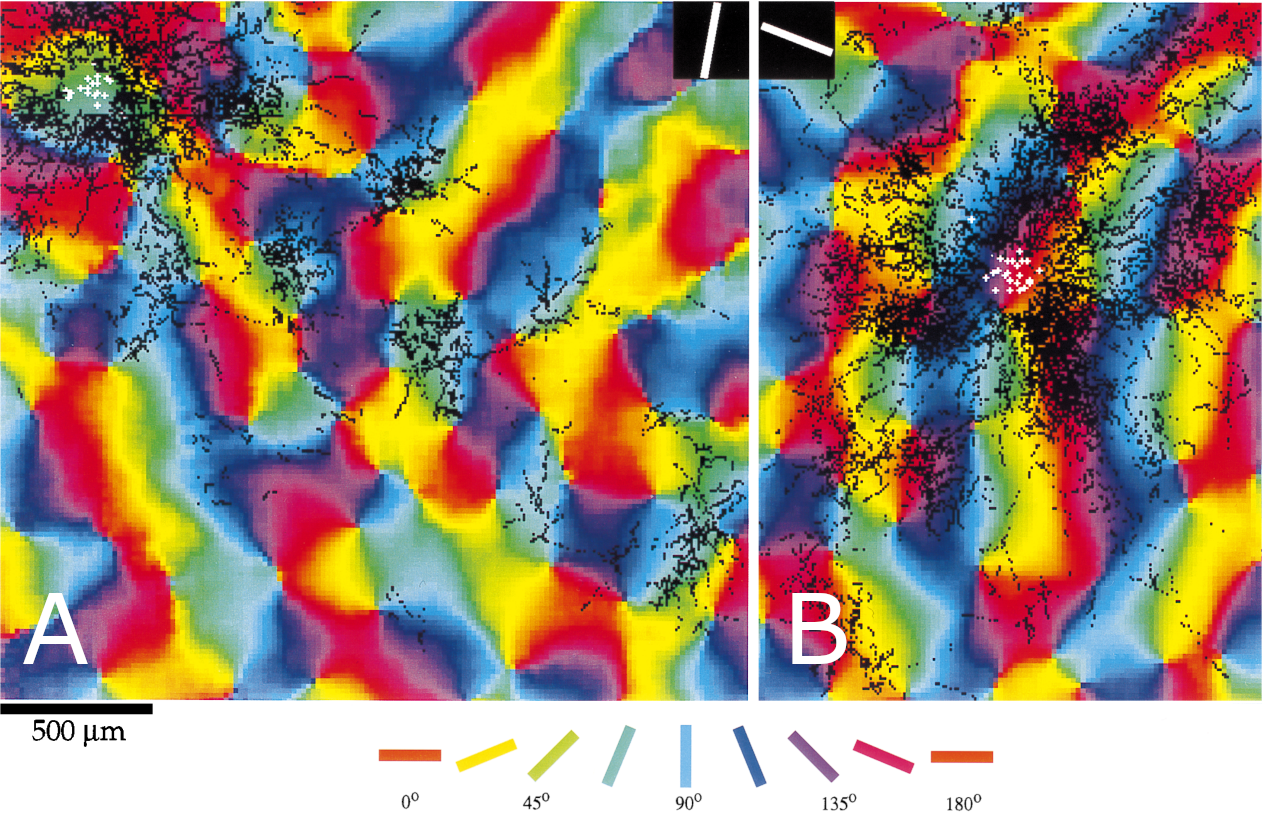}}
\caption{
\label{f:bosking} 
Lateral connections in the primary visual cortex (V1) of the tree shrew. The color indicates the orientation preference of the neurons, measured through optical imaging. In A and B, the anterograde tracer biocytin was injected in the neurons marked white, and their projections are shown as black. In both, we can see that the source and the target region have the same orientation preference (in A, cyan-green, and in B, red). Adapted from \protect\cite{bosking1997orientation}.
% Articles in J of Neurosci prior to 2014 are under a permissive license 
}
\end{figure}

So far, we saw that the original and modern CNN variants faithfully incorporate the afferent connections (input to the first conv layer) and long-range projections (one conv layer to the next conv layer). However, there is yet another kind of connection that is not included in current CNN architectures: the lateral (or horizontal) connections \cite{gilbert1990lateral}, connections {\em within} a specific cortical area (Fig.\ \ref{f:bosking}). In a sense, these lateral connection are like connections within and across featuremaps in the same convolutional layer in CNN. If such connections are implemented, what computational role could they play? 

In this paper, we propose to answer the above question, by incorporating lateral connections into the CNN architecture, and test the performance and analyze the response and connection properties. This is the main novelty of our paper. Specifically, we will focus on two properties of lateral connections: (1) recurrent activation, and (2) separation of excitatory and inhibitory connections. For (1), we will model the lateral connections among the featuremaps within the same convolutional layer using shared afferent weights and unfolding through shared lateral weights, and for (2), we will design custom loss functions to force excitatory or inhibitory weights over different lateral connection bundles. Our experiments with four benchmark data sets show improved performance, and activation and connection properties similar to those found in the biological counterpart.

\section{Background and Related Works} \label{section: bgknowledge}

The first lateral connection property we will consider is recurrent
activation through these connections.%
\commentout{One potential role of the lateral connections in V1
is that their response may become more focused through multiple
loops. \cite{miikkulainen2006computational} suggest that through this,
redundant image information is reduced through \textit{decorrelation}
\cite{barlow1989unsupervised}. This behavior was similarly observed in a
laterally connected V1 model using the NIST handwritten digits dataset
\cite{miikkulainen2006computational,wilkinson1993machine}. Since our
first CNN architecture implements lateral connections using recurrent
connections, we anticipate that similar phenomena will emerge.
} 
\textcolor{black}{There is an extensive body of research on
the role of recurrent connections in visual processing
\cite{kar2019evidence,linsley2020recurrent,kubilius2019brain}.
Furthermore, studies have shown that incorporating such biologically motivated
recurrence into CNNs often leads to performance improvements over
feedforward models \cite{liang2015recurrent,spoerer2017recurrent}. 
However, these models did not treat the recurrent connections in the context of lateral connections within each visual cortical area, thus they missed the opportunity to draw parallels with the rich response properties and connection properties found in laterally connected, biologically motivated visual cortical models (e.g.\ \cite{miikkulainen2006computational}). For example, these properties include sparsification of neural response through successive recurrent activation, and the specificity of lateral connections preferring neurons with similar orientation preference (Fig.\ \ref{f:bosking}).
}
\par
\textcolor{black}{
The second lateral connection property to be investigated is inspired by the
separate excitatory and inhibitory connections found in the lateral
connections of V1. The study of excitatory and inhibitory
connections began in the early days of neuroscience, starting with Dale's law.
Dale's law states that each neuron can only secrete one type of neurotransmitter, thus it can only be excitatory (glutamate) or inhibitory (GABA) but not both \cite{dale1935pharmacology}. %\cite{van1996chaos,brunel2000dynamics}. 
  Recently, artificial neural networks complying with Dale’s law have been proposed with various architectures (feedforward, recurrent, and convolutional)
\cite{li2024learning,cornford2020learning,blauch2022connectivity,xiao2018biologically,liao2016important}. 
However, these models employ hard constraints to enforce
Dale’s law: strictly positive and negative weight matrices
\cite{li2024learning}, strictly non-negative synaptic weights
\cite{cornford2020learning}, dedicated excitatory/inhibitory
outgoing sheets combined with layer normalization for stability
\cite{blauch2022connectivity}, and sign-constraints on weight matrices
\cite{xiao2018biologically,liao2016important}. On the contrary, our model
differs in several key aspects: (1) it is loosely inspired by Dale's law: it does not impose hard constraints on weight matrices, (2) it requires no normalization (e.g., batch or layer normalization), (3) it is placed in the context of lateral connectivity, and (4) it employs novel custom loss functions to study the emergence of excitatory and inhibitory constraints. Also, separating excitatory and inhibitory connections like this can help make gradient-based methods more biologically plausible, by solving the sign-transport problem \cite{liao2016important}.
}
\section{Methods}
\textcolor{black}{We conducted two experiments to investigate lateral
connection mechanisms. The first model focuses on recurrent
activation, analyzing the response properties of lateral connections,
the organization of lateral weights, and their relation to afferent
weights. The second model examines the effects of excitatory and
inhibitory lateral connections.}
%Below, we will provide details on the two lateral connection properties to be incorporated into CNN: (1) recurrent activation, and (2) excitatory/inhibitory separation.

\subsection{Model 1: Recurrent Activation in Laterally Connected CNN (LC-CNN)}

\begin{figure}[!h]
    \centering
    \includegraphics[width=0.8\columnwidth]{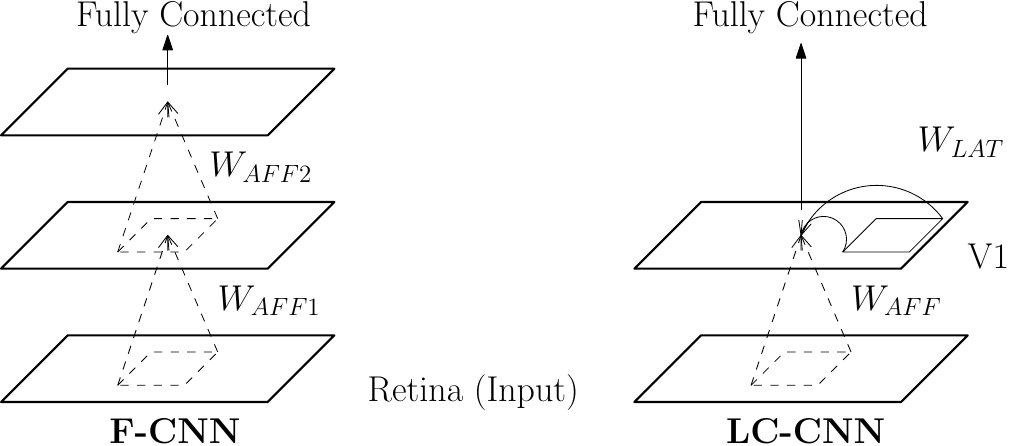}
    \caption{Left: F-CNN (baseline). $W_{\textit{AFF1}}$ and $W_{\textit{AFF2}}$ refer to the first and subsequent afferent convolution weights. Right: Laterally Connected-CNN. $W_{\textit{AFF}}$ and $W_{\textit{LAT}}$ indicate the afferent and lateral convolution weights, respectively. Both models go through convolution through two sets of weights.}
    \label{fig: exp1_0}
\end{figure}

The CNN architecture in our first experiment is designed to test the
effect of lateral connections within a specific visual area
(e.g., V1) and recurrent activation through these connections. For simplicity, 
we bypass the lateral geniculate nucleus (LGN), thus we only have the retina (the input image)  and the V1 layer. Fig.\ \ref{fig: exp1_0} shows the
design of our first model LC-CNN and the feedforward CNN (F-CNN) baseline.

%In the LC-CNN, an image $I$ is convolved with $W_{\textit{AFF}}$. Subsequently, a single set of $W_{\textit{AFF}}$ and $W_{\textit{LAT}}$ is \textcolor{black}{used and re-used} repeatedly to generate successive output by feeding through itself multiple times, before the final output is sent to the fully connected layer.  
For F-CNN, Eq.\ \ref{eq: exp0-1} below shows the computations leading to the 
last conv layer output $O_{\textit{AFF2}}$. \textit{R$(\cdot)$}
is the ReLU activation function, and * the convolution
operator. Here, an input image $I$ is convolved with
$W_{\textit{AFF1}}$. After that, the previous layer's output is convolved with $W_{\textit{AFF2}}$, passing its output to the fully connected layer.
\begin{equation}
    \label{eq: exp0-1}
    O_{\textit{AFF2}} = \textit{R}\left(W_{\textit{AFF2}} \conv \textit{R}\left(W_{\textit{AFF1}} \conv I\right)\right)
\end{equation}

\begin{figure}[!tbp]
    \centering
    \includegraphics[width=0.8\columnwidth]{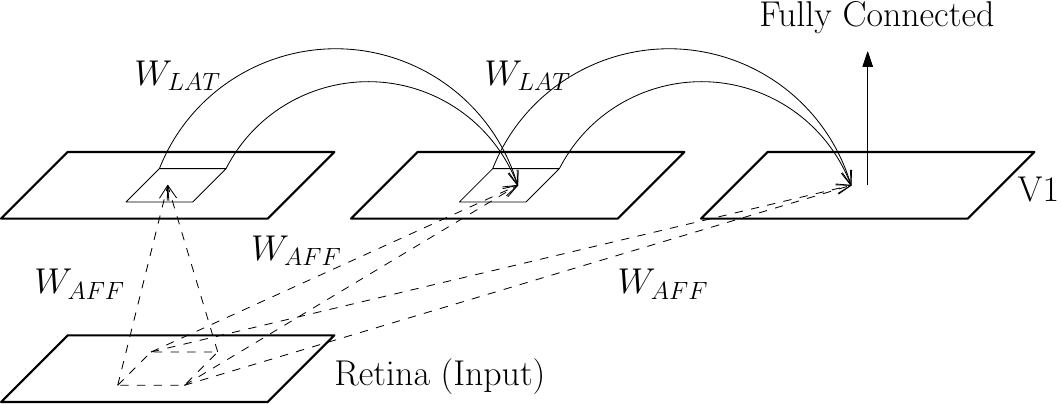}
    \caption{LC-CNN with one loop unrolled (LC-CNN: Loop-1). The afferent activation is computed once in the beginning and reused ($W_{\textit{AFF}}$ are shared). $W_{\textit{LAT}}$ are shared throughout all loops. The input and output of $W_{\textit{LAT}}$ have the same channel depth because it loops back into itself. $I$ is an input image. Note that the model only uses two sets of convolutional weights ($W_{\textit{AFF}}$ and $W_{\textit{LAT}}$). Note that RCNN has the same architecture \protect\cite{liang2015recurrent}, but with a different interpretation.}
    \label{fig: exp1}
\end{figure}

LC-CNN includes recurrent activation, so we need to unroll it. Fig.\ \ref{fig: exp1} shows how this is done. The process first performs convolution $W_{\textit{AFF}}$ with an input image $I$. Next, $O_{\textit{LAT}}$ that represents a V1 sheet generates activation using both $O_{\textit{AFF}}$ and the previous output of $O_{\textit{LAT}}$, enabling the simultaneous learning of $W_{\textit{AFF}}$ and $W_{\textit{LAT}}$. For this to work, the input and output channel sizes should match in $O_{\textit{LAT}}$. In other words, the output sizes of $O_{\textit{AFF}}$ and $O_{\textit{LAT}}$ should be the same. Note that both $W_{\textit{AFF}}$ and $W_{\textit{LAT}}$ are shared across all time steps $t$.  
Eq.~\ref{eq: exp1-1}-\ref{eq: exp1-3} below summarize these steps. 
\begin{align}
    \label{eq: exp1-1}
    O_{\textit{AFF}} &= \textit{R}\left(W_{\textit{AFF}} \conv I\right)
    \\
    \label{eq: exp1-3}
    O_{\textit{LAT}}(t) &= \textit{R}\bigg(W_{\textit{LAT}}\big(
    O_{\textit{AFF}} \oplus O_{\textit{LAT}}(t-1)\big)\bigg), \nonumber \\
    &\text{for } t \geq 0 \text{ with } O_{\textit{LAT}}(-1) = 0
\end{align}
where $\oplus$ is the element-wise addition operator. \textcolor{black}{(Note that the number of parameters in F-CNN and LC-CNN are equal due to the weight sharing in LC-CNN.)} Back Propagation Through Time (BPTT) \cite{werbos1990backpropagation} is used for training, with the following loss function:
\begin{align} \label{eq: lossfunction}
     L = CE\left(Y_{\textit{true}}, Y_{\textit{expected}}\right) + \lambda_1 \sum_{i} \lvert w_i \rvert, 
\end{align} where CE is the cross entropy, $Y_{\textit{true}}$ the ground truth, $Y_{\textit{expected}}$ the model prediction, $w_i$ the weights, and $\lambda_1$ the L1 regularization hyperparameter.
 Gradients are accumulated over all time steps, and the shared weight $W_{\textit{LAT}}$ is updated by summing these accumulated gradients. Since $W_{\textit{AFF}}$ is only used once (Eq.~\ref{eq: exp1-1}) and its output reused (Eq.~\ref{eq: exp1-3}), its gradients are not accumulated. 
See section~\ref{s:train} for training details, including data sets used.

\subsection{Model 2: Excitatory and Inhibitory Separation in Laterally Connected CNN (LCEI-CNN)}

\begin{figure}[h]
    \centering
    \includegraphics[width=1.0\columnwidth]{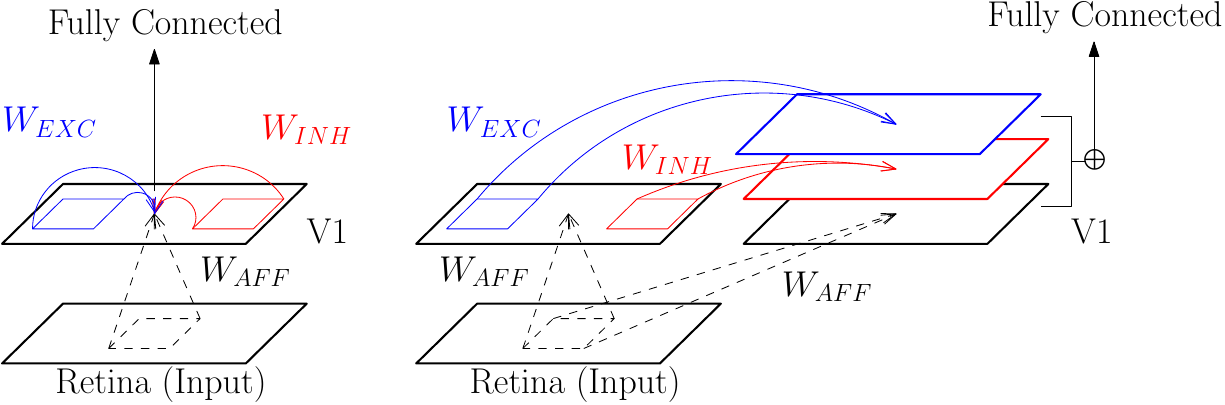}
    \caption{Lateral connections considering excitatory and inhibitory connections. This is our proposed model, LCEI-CNN. $W_{\textit{EXC}}$ and $W_{\textit{INH}}$ denote excitatory and inhibitory weights, respectively. $I$ is an input image. $W_{\textit{AFF}}$ is the afferent convolution weights before separation. See the diagram on the right in Fig.~\ref{fig: exp1_0} for the state before separation, where lateral connections are defined without distinguishing between excitatory and inhibitory connections.}
    \label{fig: exp2}
\end{figure}

The second experiment is designed to analyze the effects of separating
lateral excitatory and inhibitory connections. In the cortex, lateral interactions are either excitatory or inhibitory, as shown in Fig.\ \ref{fig: exp2} (left), due to the separate excitatory and inhibitory populations of neurons.
To model this in the existing CNN framework, we propose to form two separate paths, one with mostly excitatory weights and the other with mostly inhibitory weights (Fig.\ \ref{fig: exp2}, right). We use custom loss functions to train the respective weights to be mostly excitatory or inhibitory. We call this LCEI-CNN (excitatory/inhibitory CNN). 
%Our model incorporates a custom loss function, which will likely assign one group of weights in a positive region and the other in a negative region during and after training, leading to mostly positive activations from one pathway and negative activations from the other. This aligns with excitatory and inhibitory lateral connections in the V1 (Fig.\ \ref{fig: exp2}, left). The weights in both groups start with small random values near zero
% \commentout{\bf yschoe: line 124 : zero value or zero mean?} and gradually diverge during training, driven by the loss function. The resulting model is called ``EI-CNN'' (excitatory/inhibitory CNN).
% yschoe
% jhpark: deleted. This corresponds to Loop-1 in the previous experiment, where the excitatory and inhibitory connections are separated.}  
%Fig.\ \ref{fig: exp2} compares two different types of CNN architectures: (1) Before separating excitatory and inhibitory weights (LC-CNN) and (2) After separation (EI-CNN). 
Eq.\ \ref{eq: separation1}-\ref{eq: separation2} describe how LCEI-CNN is activated.
\par
The activation of the neurons in the model is done as follows:
\begin{align} \label{eq: separation1}
    O_{\textit{AFF}} &= \textit{R}\left(W_{\textit{AFF}} \conv I\right) \\[5pt]
    \label{eq: separation2}
    O_{\textit{LAT}} &= \sigma\left(W_{\textit{EXC}} \conv O_{\textit{AFF}}\right) \oplus \sigma\left(W_{\textit{INH}} \conv O_{\textit{AFF}}\right) \oplus O_{\textit{AFF}} 
\end{align}
where $\sigma(\cdot)$ is the tanh activation function. 
It begins with an input image \textit{I} undergoing a convolution with $W_{\textit{AFF}}$, resulting in the initial afferent activation $O_{\textit{AFF}}$. 
Subsequently, $W_{\textit{EXC}}$ and $W_{\textit{INH}}$, the excitatory and inhibitory weights, are applied to compute the lateral interactions on $O_{\textit{AFF}}$. 
% \textcolor{black}{Note that $O_{\textit{AFF}}$ bypasses $W_{\textit{EXC}}$ and $W_{\textit{INH}}$ and flows directly to $O_{\textit{LAT}}$.}
An element-wise addition follows this, passing the sum to the fully connected layer. We utilize the tanh activation function to enable $O_{\textit{EXC}}$ and $O_{\textit{INH}}$ to output positive and negative activations, respectively. 
When one set of weights resides in the positive region, the corresponding output is a positive feature map; conversely, when the other set of weights is in the negative region, it outputs a negative feature map. 
This implements the effect of separate excitatory and inhibitory lateral contributions. 
Custom penalty terms are then used to encourage weight differentiation during the LCEI-CNN training. 
\begin{table*}[!th]
  \centering
  \begin{tabular*}{\textwidth}{@{\extracolsep{\fill}}lcccc@{}}
    \toprule
    Structure / Dataset & MNIST & Fashion-MNIST & CIFAR-10 & Natural Images \\
    \toprule
    (Baseline) F-CNN       & 97.38 $\pm$ 0.36 & 87.64 $\pm$ 0.39 & 49.95 $\pm$ 2.36 & 79.02 $\pm$ 0.75\\
    \midrule
    (Ours) LC-CNN: Loop-1        & 98.04 $\pm$ 0.09 & 88.50 $\pm$ 0.45 & 56.26 $\pm$ 1.75 & 79.85 $\pm$ 1.17\\
    (Ours) LC-CNN: Loop-3        & 98.38 $\pm$ 0.08 & 89.30 $\pm$ 0.23 & 58.09 $\pm$ 1.35 & \textbf{81.58 $\pm$ 0.75}\\
    (Ours) LC-CNN: Loop-5        & \textbf{98.52 $\pm$ 0.18} & \textbf{90.04 $\pm$ 0.05} & \textbf{58.62 $\pm$ 0.23} & 80.03 $\pm$ 1.02\\
    \bottomrule
  \end{tabular*}
  \caption{Comparison of test accuracy between the F-CNN, LC-CNN: Loop-1, LC-CNN: Loop-3, and LC-CNN: Loop-5 designs. For each experiment, the mean test accuracy and its standard deviation are provided across five runs. The best accuracy is shown in bold.}
  \label{table: exp1}
\end{table*}
\par
To enable excitatory and inhibitory connections, we developed a penalty term that can be included in the loss function. The penalty term we've developed is implemented similarly to well-known regularization techniques such as L1 and L2 (lasso and ridge regression) \cite{l1tibshirani1996regression,l2hoerl1970ridge}. This allows for its effective combination with the cross-entropy loss \cite{zhang2018generalized}. We devised four different penalty terms.
\par
In our first approach to the penalty term for weight separation (Eq.\ \ref{eq: ws1}), we start by calculating the mean of $W_{\textit{EXC}}$ and $W_{\textit{INH}}$. We then compute the absolute difference between these two values and subsequently negate it. \textcolor{black}{Note that either one of $W_{\textit{EXC}}$ or $W_{\textit{INH}}$ can become positive, but always, the other one will become negative. We can simply relabel EXC and INH in case there is a mismatch in the resulting sign.}
% yschoe:  It is important to note that while $W_{\textit{EXC}}$ and $W_{\textit{INH}}$ can be interchanged in this process, it does not alter the function's essence due to the use of the absolute value.
\begin{align} \label{eq: ws1}
    L_{\textit{ABS}} &= -\lvert\mathbb{E}[W_{\textit{EXC}}] - \mathbb{E}[W_{\textit{INH}}]\rvert
\end{align}
An issue with the first approach is that both weights will grow without bounds.\commentout{What do you mean by "saturate"? You mean "grow without bound"?} In other words, when this penalty is added to the loss, there is a possibility that $W_{\textit{EXC}}$ and $W_{\textit{INH}}$ may perpetually diverge from each other. This ongoing separation could cause the model to prioritize driving these two weights apart without sufficiently accounting for the cross-entropy loss. Therefore, we introduce a saturation mechanism using the tanh function $\sigma(\cdot)$ (Eq.\ \ref{eq: ws2}). Again, the label EXC/INH can be interchanged based on the outcome.
\begin{align}
    \label{eq: ws2}
    L_{\textit{SAT-ABS}} &= -\lvert\sigma\left(\mathbb{E}[W_{\textit{EXC}}]\right) - \sigma\left(\mathbb{E}[W_{\textit{INH}}]\right)\rvert
\end{align}
For the third approach, we dropped the absolute value function in our penalty term (Eq.\ \ref{eq: ws3}). Instead of focusing on the difference that drives the weights apart, our redesigned penalty term allows each weight to establish weight separation independently. It should be noted that the weight labels (EXH/INH) cannot be interchanged since we apply negation to $W_{\textit{EXC}}$, and we dropped the absolute value function.
\begin{align} \label{eq: ws3}
     L_{\textit{EXP-SAT}} &= \sigma\left(\EX[-W_{\textit{EXC}}]\right) + \sigma\left(\EX[W_{\textit{INH}}]\right)
\end{align}
In the fourth case (Eq.\ \ref{eq: ws4}), we switched the order of applying the expected value operator and the tanh function because their sequence may impact weights during training.
\begin{align} \label{eq: ws4}
     L_{\textit{SAT-EXP}} &= \EX[\sigma\left(-W_{\textit{EXC}}\right)] + \EX[\sigma\left(W_{\textit{INH}}\right)]
\end{align}
We incorporated these penalty terms into the loss function, enabling us to train the model with a loss function that integrates cross-entropy (CE), L1 regularization (hyperparameter = $\lambda_1$), and a choice of weight separation loss $L_{ws}$, where $ws \in \{\textit{ABS, SAT-ABS, EXP-SAT, SAT-EXP}\}$. To tune the strength of the penalty, we introduced hyperparameter $\lambda_2$ and tested it with search space $[0, 0.1, 1, 2.5, 5, 7.5, 10]$ (See Eq.\ \ref{eq: lossfunction with pen}). By adopting this loss function, the model's weights, $W_{\textit{EXC}}$ and $W_{\textit{INH}}$, initially close to zero, will diverge during training.
\begin{align} \label{eq: lossfunction with pen}
     L = CE\left(Y_{\textit{true}}, Y_{\textit{expected}}\right) + \lambda_1 \sum_{i} \lvert w_i \rvert + \lambda_2 L_{ws},
\end{align}
where $Y_{true}$ is the ground truth, $Y_{expected}$ is the network's prediction, and $w$ is the network's weights.
\par
\textcolor{black}{We note here that this implementation does not
strictly adhere to Dale’s principle. Our aim was to preserve the original CNN architecture, without putting hard constraints like this. We believe the use of a penalty term in the loss function allows us to observe better the functional significance of excitatory-inhibitory separation. If this kind of separation did not have any functional significance in the CNN, we would not observe any such separation.} Section~\ref{s:train} will discuss training details.
% \textcolor{black}{In addition, our method allows an individual neuron to have any signs of output weights (i.e., we do not apply sign-transport or strict measures to limit a neuron's behavior - we use a custom loss function to induce such behavior \textit{naturally}). This flexibility reflects the early neural developmental phase when individual neurons may release both excitatory and inhibitory neurotransmitters \cite{?}. However, as neural circuits mature, neurons typically specialize in either excitatory or inhibitory transmission, following Dale's principle by releasing only one type of neurotransmitter. Initially, in this experiment, weight values are randomly distributed before training analogous to pre-maturation. After training, similar to post-maturation, excitatory and inhibitory weights increasingly localize to positive and negative regions, respectively.}
\subsection{Training Details} 

\label{s:train}

%yschoe : I think this is distracting. 
%Growing evidence supports the theory that the visual system uses a sparse coding mechanism to represent visual stimuli, where information is efficiently encoded by a limited number of cells \cite{lian2019toward}. Following this, in our experiments, we applied L1 regularization \cite{l1tibshirani1996regression} to create a sparse and selective environment inspired by the brain's efficiency in processing information sparsely. 
%L1 regularization is well known for its ability to induce sparsity in the coefficients (i.e., weights) \cite{lee2006efficient, ng2004feature}. This behavior effectively leads to the selective activation of neurons in neural networks. This aligns with our goal of developing the biologically motivated CNNs. We set the L1 regularization strength to 1e-3 across all experiments, as it effectively preserves test accuracy and promotes weight sparsity.
All convolutional layers had 8 channels for model 1 and 4 channels for model 2. The receptive field size was $7 \times 7$ for all convolutions. Intel i9-13900HX CPU and an RTX 4070 laptop GPU were used for training (1 to 2 hours for model 1, and 3 to 4 hours for model 2). See SM \ref{s:archi} and SM \ref{s:compute} for more CNN architecture and computing resources details.
\par For all experiments, we used L1 regularization on the weights \cite{l1tibshirani1996regression,lee2006efficient} ($\lambda_1$=1e-3 across all experiments). In our first experiment, we utilized Stochastic Gradient Descent (SGD) with momentum \cite{ruder2016overview} value of $0.9$ and the learning rate tuning with 1e-2, 1e-3, and 1e-4. For the second experiment, we used the Adam optimizer \cite{kingma2014adam}, using the same range of learning rates. In both experiments, we avoided using extra computational processes such as batch normalization \cite{ioffe2015batch} or local response normalization \cite{krizhevsky2017imagenet} in the post-processing stage of the convolutional filters. This approach allowed us to focus solely on the impact of the new structures and avoid potential confounding effects from the additional computations. 
\par
In all experiments, the models used four benchmark datasets: MNIST \cite{deng2012mnist}, Fashion-MNIST \cite{xiao2017fashion}, CIFAR-10 \cite{krizhevsky2009learning}, and Natural Images \cite{roy1807effects}. Training/validation sets were 85\% and 15\% of 60k, 60k, 50k, and 6k samples; and test sets were 10k, 10k, 10k, and 800 samples, respectively. In all cases, images were gray-scaled, resized to 48$\times$48, and  kernel size 7$\times$7. This choice of larger kernel size (usually 3$\times$3), was deliberate because we wanted to compare the convolution kernels to those observable in the visual receptive fields of the V1 and the lateral connection patterns \cite{krizhevsky2017imagenet,jones1987evaluation}. % This larger kernel size allows for better analysis of these patterns, providing insights that might be less discernible with smaller kernels (e.g., 3$\times$3 or 5$\times$5). 

\section{Experiments and Results}
We tested the two laterally connected CNN models in terms of (1) performance compared to baseline and (2) analysis of connection weight and neural response (activation) properties, compared to known results in neuroscience.
\subsection{Model 1: Recurrent Activation of Laterally Connected CNN (LC-CNN)}
\noindent
\paragraph{Performance:}
We evaluated the performance between the baseline F-CNN (the vanilla CNN) and LC-CNN with one, three, and five loops (Loop-1, Loop-3, Loop-5) across four datasets, maintaining the same number of parameters across all experiments. LC-CNN demonstrated better test accuracies than F-CNN for all structures and datasets, as shown in Table \ref{table: exp1}. It can be observed that LC-CNN tends to perform better as the number of lateral loops increases. \textcolor{black}{This is not something that is unexpected, since it is well known that deeper CNNs with more layers tend to perform better, and more unrolled loops are equivalent to deeper layers. However, in our case, all models had the same number of tunable parameters through shared weights.}
%\textcolor{black}{Focusing solely on model structure, laterally or recurrently connected CNNs (i.e., LC-CNN) can be viewed as simply increasing the depth of the network, which typically leads to better performance compared to feedforward CNNs (i.e., F-CNN). This improvement is expected, as laterally connected CNNs can be unfolded over time and represented as deeper feedforward CNNs (See Fig.\ \ref{fig: exp1}). For instance, it can be thought that a Loop-\textit{N} LC-CNN is equivalent to a \textit{N}+2 F-CNN with shared weights across layers. \textit{N}+2 comes from the recurrent iterations \textit{N} plus the original number of layers 2.}
 \textcolor{black}{Also note that the performance overall may not be very high compared to the state of the art, since we are using a very minimal, restrictive CNN architecture in order to directly assess the impact of lateral connections.}
\noindent
\paragraph{Analysis (Neural Activation):}
\par
\begin{figure}[!htbp]
    \centering
    \includegraphics[width=0.8\columnwidth]{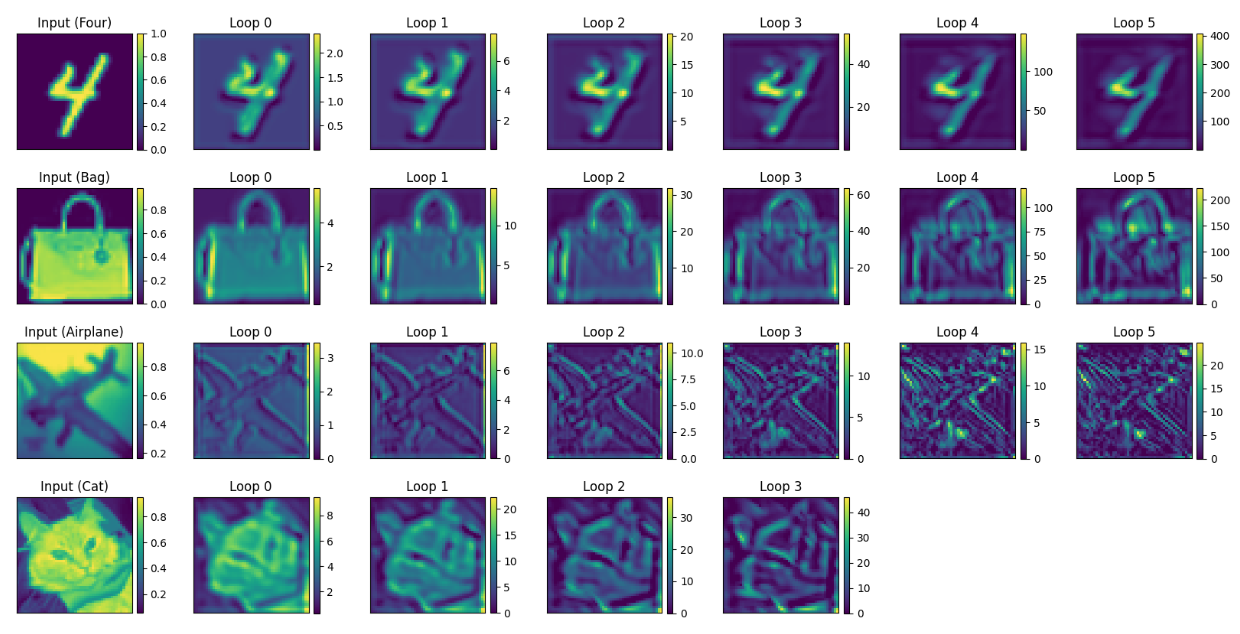}
    \caption{Changes in featuremap $O_{\textit{LAT}}$ over lateral activation loops. The first column is the input, and the second to the last column show 0 to 5 loops (0 loop is equivalent to FCNN). Top to bottom: MNIST, Fashion-MNIST, CIFAR-10, and Natural Images. Each response image is the sum of featuremaps in all channels. }
    \label{fig: exp1_loop}
\end{figure}
\begin{figure}[!htbp]
    \centering
    \begin{subfigure}{0.4\columnwidth}
        \centering
        \includegraphics[width=\linewidth]{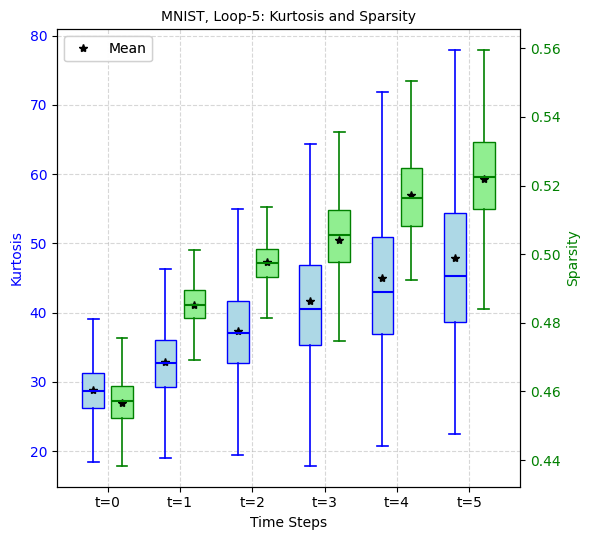}
    \end{subfigure}
    \hspace{0.02\columnwidth} % Adjusts horizontal spacing between columns
    \begin{subfigure}{0.4\columnwidth}
        \centering
        \includegraphics[width=\linewidth]{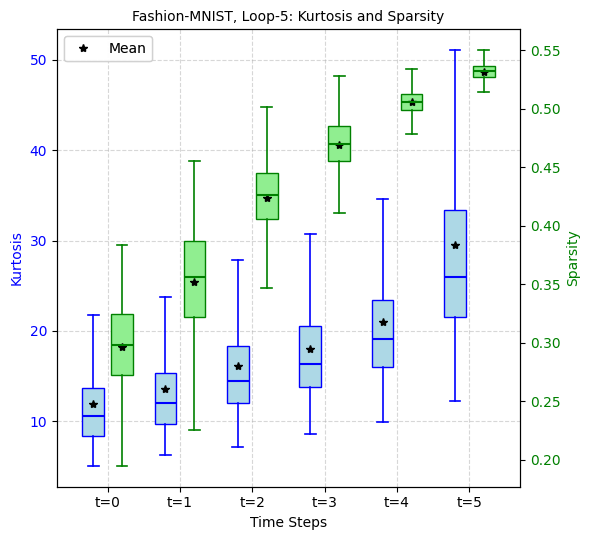}
    \end{subfigure}
    \begin{subfigure}{0.4\columnwidth}
        \centering
        \includegraphics[width=\linewidth]{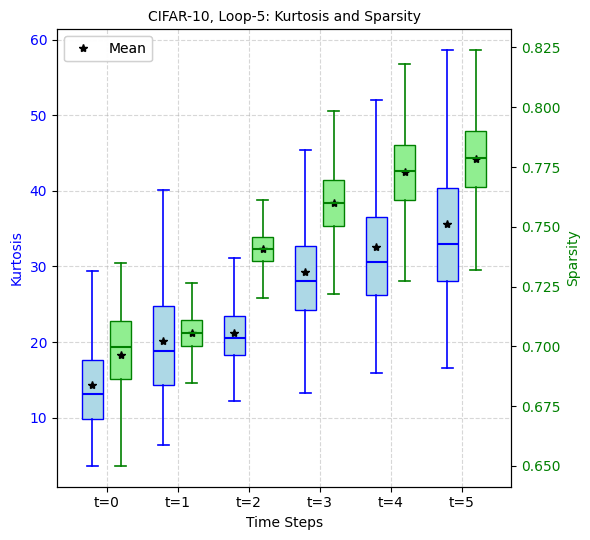}
    \end{subfigure}
    \hspace{0.02\columnwidth} % Adjusts horizontal spacing between columns
    \begin{subfigure}{0.40\columnwidth}
        \centering
        \includegraphics[width=\linewidth]{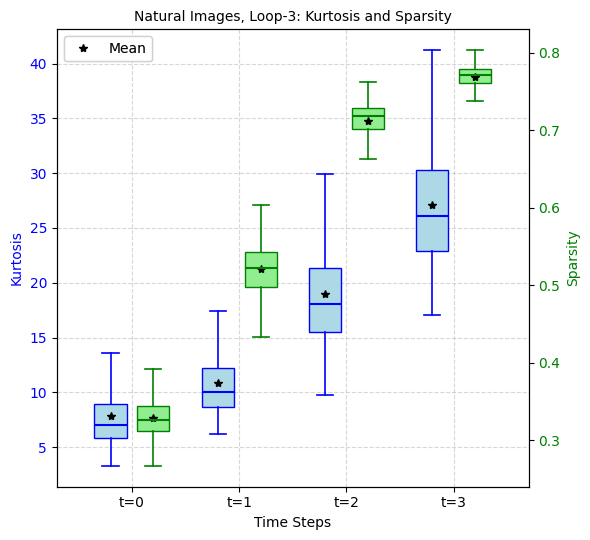}
    \end{subfigure}
    \caption{Kurtosis and sparsity in $O_{\textit{LAT}}$. The kurtosis and sparsity are measured in the response to each image, and the mean, median, and standard deviation of all responses computed. Results are from best test accuracy trials. The kurtosis and sparsity increases as the loops are increased.}
    \label{fig: exp1_kurtosis}
\end{figure}

One of the main purposes of this paper is to analyze the neural activation properties of the laterally connected CNN with its biological counterpart, the mammalian primary visual cortex (V1). Our first step is to observe the response in the featuremaps (FMs) over increasing number of lateral activation loops.
Fig.\ \ref{fig: exp1_loop} shows the FM activation ($O_{\textit{LAT}}$) over the loops. We can see that the background becomes darker and the foreground brighter, meaning that the response becomes sparser. Fig.\ \ref{fig: exp1_kurtosis} shows this trend quantitatively. Kurtosis (fourth central moment) is a well-known measure of sparsity \cite{barlow1972single}, and sparsity can also be measured directly by counting zero values in FM activation.

Sparsity in cortical activation has been theorized as playing an important role in neural coding and decorrelation, and actual evidence has been found in the primary visual cortex (V1) \cite{olshausen1996emergence,vinje2000sparse,la2021finer}. In terms of computational models, \cite{miikkulainen2006computational} showed that a laterally interconnect self-organizing model of V1 achieves such sparseness through Hebbian learning, and recurrent activation over the lateral connections in a similar manner as we have shown. However, such a mechanism has not been used in CNNs, to our knowledge. Sparse activation is common in CNNs as higher convolutional layers are reached, but our result is interesting because this sparsity is achieved through shared lateral weights. Furthermore, sparsity emerged without any explicit loss term to enforce sparsity. There are models that utilize sparse activation, but they use an explicit sparse activation function such as top-k \cite{bizopoulos2020sparsely}. Models such as sparse SNN \cite{liu2015sparse} and sparse spiking CNN \cite{cordone2021learning} also exist, but these models were more about sparser convolutional kernels for efficient processing. 

\commentout{
However, our focus is not on enhancing performance. Rather, we are interested in how these LC-CNNs emulate biological processes.
Our investigation revealed findings beyond the comparative analysis of test accuracies. In particular, we observed that LC-CNN exhibits characteristics akin to biological vision systems and the computational model of V1. The V1 layer activations in LC-CNN displayed increasing kurtosis and sparsity as more recurrent loops were executed. % \commentout{Moreover, the afferent and lateral weights of LC-CNN exhibited a linear relationship, providing strong evidence that the V1 layer forms orientation maps.} % yschoe: check if the statement below is okay 
Moreover, the lateral connections tend to be established between feature maps that have similar afferent weight properties (i.e., similar orientation preference). These findings suggest that the network's behavior parallels certain aspects of biological neural processing, highlighting the potential of our model to mimic visual cognitive mechanisms.
}

\begin{figure}[!htbp]
    \centering
    \includegraphics[width=0.95\columnwidth]{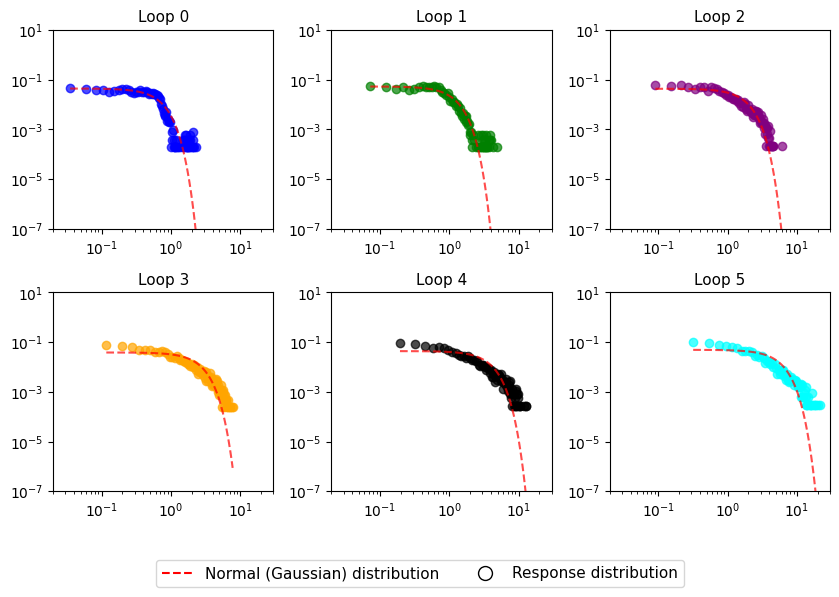}
    \caption{Response histogram from CIFAR-10. Normal distribution (scaled to match the variance) and lateral activation $O_{\textit{LAT}}$ distribution plotted on a log-log scale (for strictly positive output values only, due to the log-scale). The same input image is used from Fig.\ \ref{fig: exp1_loop}, third row. Starting from loop 3, we can see that the probability increases in the lowest and highest range, compared to the Gaussian.}
    \label{fig: exp1_cifar_powerlaw}
\end{figure}

Another way to view sparsity and its functional role is to observe the response distribution. Fig.\ \ref{fig: exp1_cifar_powerlaw} shows how the response distribution changes over the lateral activation loop (log-log plot), initially close to a normal distribution (dashed red curve), but becoming closer to a power-law  (declining linear line). It was proposed in \cite{lee2003detecting,sarma:aaai06} that the intersection of the response distribution curve and the normal distribution curve may have an important functional significance: it has a linear relationship with the perceptual threshold for salience. The models described in \cite{lee2003detecting,sarma:aaai06} was a series of convolutions to explicitly model the LGN and V1 processing with fixed kernels (difference of Gaussian and oriented Gabor patterns, respectively). It is notable that similar results can be obtained in CNN, but only when lateral interactions are used.

% yschoe-2005-08-21 : moved from the appendix

\commentout{
Another interesting property of LC-CNN that may have a counterpart
in biology is the response distribution of the feature maps. For
example, \cite{lee2003detecting} showed that a visual cortical
model based on orientation energy computed using oriented Gabor
filters based on \cite{geisler2001edge} exhibited a power-law in its response
distribution. They also observed that when the response distribution
is compared with a normal distribution with equivalent variance, the
intersections are linearly correlated with the perceptual threshold for edge salience. To test the response properties of our LC-CNN, we computed the feature
maps' response distribution and plotted them in as a log-log plot in
Fig.\ \ref{fig: exp1_cifar_powerlaw}. The plot shows
the response distribution as the number of recurrent activations is
increased. Initially, the response is similar to the normal distribution
(dashed red curve), but as the loops increase, the response becomes
closer to a power-law (straight declining line).

This serves as another interesting link between the Laterally Connected CNN and the biological visual system. Further analysis is needed to understand the functional significance of this result, e.g., whether additional tasks such as saliency detection becomes easier using a CNN with lateral activation.
}
This may have some functional significance in visual cortical processing. 
For example, it was shown that the intersection point of the power-law-like response distribution and the matching Gaussian (same variance) meets where the heavy tail begins is linearly correlated with the saliency threshold marked by humans \cite{lee2003detecting}. Furthermore, the same model, presented with white noise image, would give a near-Gaussian response \cite{sarma:aaai06}. These results point to the functional significance of the power-law-like property for down-stream tasks. These results may also have deep theoretical implications as well. Directly solving the power law = Gaussian, we get the Lambert $W$ function:
\begin{equation}
    c \frac{1}{x^{a}} = \frac{1}{\sigma\sqrt{2\pi}} e^{-\frac{x^2}{2\sigma^2}},
\end{equation}
which then gives 
\begin{equation}
    x =    \pm \sqrt{ - a\sigma^2 W\left(                                                        -\frac{(c\sigma\sqrt{2\pi})^{2/a}}{a\sigma^2} \right)},                         
\end{equation}
where $c$ is a normalization constant, $a$ is the power law exponent and $\sigma$ is the standard deviation (see \ref{s:solve}). The Lambert $W$ function is defined as  $W(z)e^{W(z)} = z$ \cite{corless1996lambert}. All three functions are ubiquitous in nature, and these results point to a deeper computational principle embodied in visual cortical processing.

\commentout{
\par
Fig.\ \ref{fig: exp1_kurtosis} illustrates that both kurtosis and sparsity increase with each successive loop in LC-CNN. We observed that this increasing trend is consistent across all other structures. These observations were derived from the testing dataset, and we confirmed negligible differences between kurtosis and sparsity of training and testing data. Kurtosis is higher when the distribution is peaked at zero, meaning a larger number of neurons are inactive, compared to the normal distribution. It is well-established that the sparsity of neural activity can be quantitatively measured using kurtosis \cite{barlow1972single,field1994goal}. We also directly computed sparsity by counting the number of zero-values in the CNN activation. The observed increase in kurtosis is similar to those observed in a biologically plausible V1 model LISSOM \cite{miikkulainen2006computational}.
}

\begin{figure}[!htbp]
    \centering    \includegraphics[width=\columnwidth]{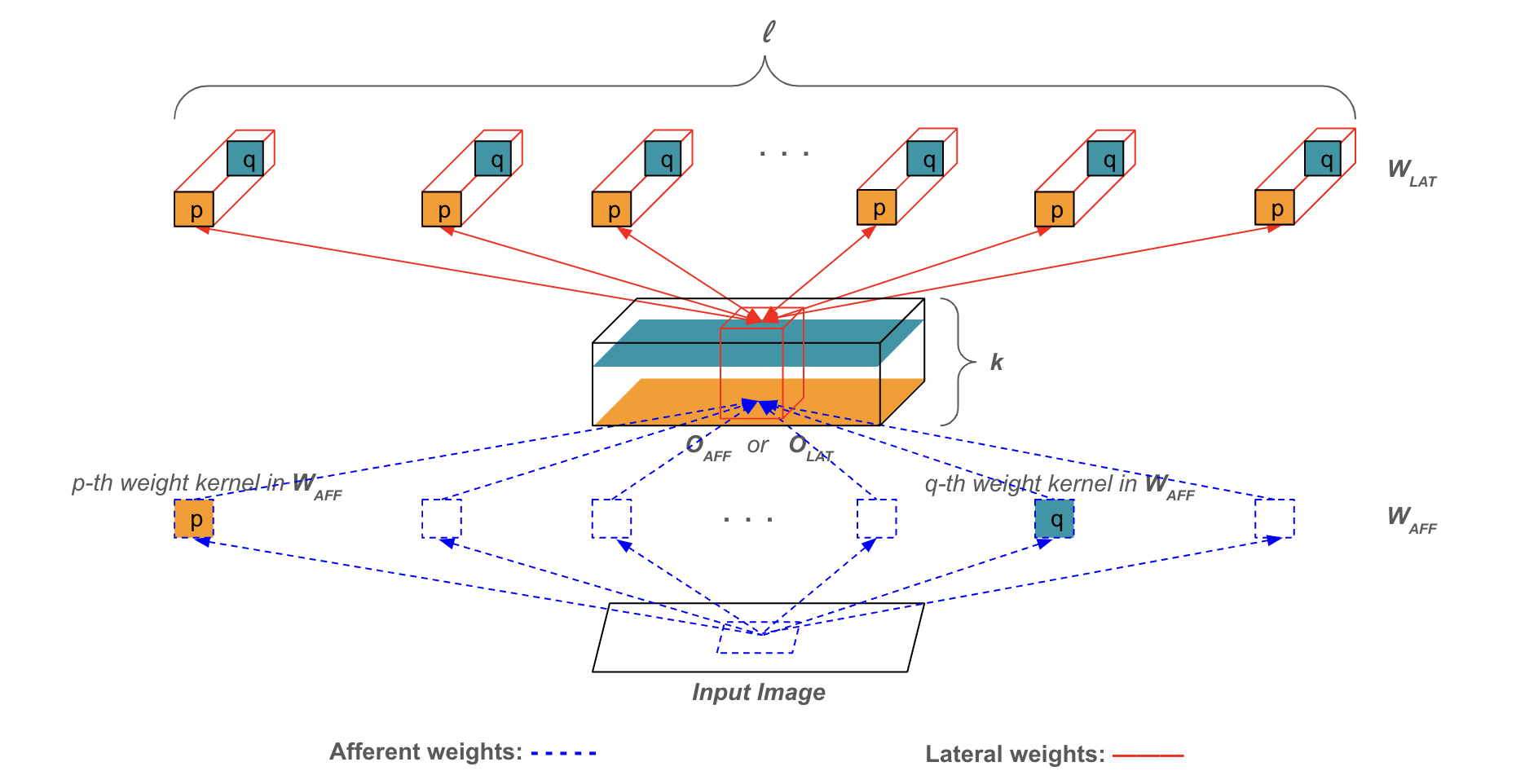}
    \caption{Comparison of afferent ($W_{\textit{AFF}}$) and lateral weight ($W_{\textit{LAT}}$) properties in LC-CNN. $k$ feature maps are generated by $W_{\textit{AFF}}$. This results in $O_{\textit{AFF}}$. Similarly, $l$ feature maps are generated by $W_{\textit{LAT}}$. This results in $O_{\textit{LAT}}$. By comparing two similarity pair values: (1) $p$-th and $q$-th afferent weight kernel in $W_{\textit{AFF}}$ and (2) $p$-th and $q$-th lateral weight kernel in $r$-th tensor in $W_{\textit{LAT}}$, where $0 \leq r \leq l$, we can compare the similarity between $W_{\textit{AFF}}$ and $W_{\textit{LAT}}$.}
    \label{fig: weight sim des}
\end{figure}

\paragraph{Analysis (Connection Properties):}
As shown in Fig.\ \ref{f:bosking}, the lateral connections in the biological visual cortex have the propensity to connect regions that have similar orientation preference. This kind of arrangement is theorized to provide the anatomical basis for contour detection \cite{miikkulainen2006computational,geisler2001edge}. There is a challenge though, since the lateral connections ($W_{\textit{LAT}}$) in our model cannot be directly mapped to the biological counterpart due to CNNs using the convolution operation. In CNN, each channel forms its own feature map, thus each feature map in its entirely only has a single afferent feature represented, where as in the visual cortex, the single sheet contains a patchwork of orientation preferences as in Fig.\ \ref{f:bosking}. However, we can still examine the relationship between afferent and lateral connections.

\commentout{
On the other hand, the lateral connection weights in LC-CNN exhibited patchy connections (Fig.\ \ref{fig: exp1_weight kernel}). These results are in line with the findings in V1 by \cite{bosking1997orientation}. 
\textcolor{black}{Furthermore, Bosking et al. showed that these patches tend to link between cortical columns (collections of neurons arranged in vertical columns that share similar stimulus specificity) that have similar orientation preferences.}
% jhpark dramatically incorrect sentence: Furthermore, Bosking et al.\ showed that these patches tend to link between cortical columns (a collection of neurons arranged in a vertical column that have similar stimulus specificity) that have similar orientation preferences.
It is hard to directly compare this to our results since, strictly speaking, in a single convolution layer of LC-CNN, each feature map from one channel roughly corresponds to one orientation column in V1. However, we can check if the slices of $W_{\textit{LAT}}$ that connect one feature map to another exhibit a similar property, 
\textcolor{black}{treating each feature map slice as an orientation column}. % jhpark: removed brackets
}
\par

\begin{figure}[!htbp]
\begin{tabular}{@{}c@{}c@{}c@{}c@{}c@{}}
\rotatebox{90}{~~~~~~{\normalsize before}} & 
        \includegraphics[width=0.225\columnwidth]{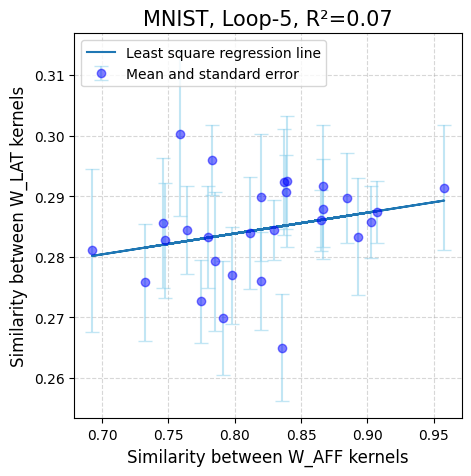} &
~
        \includegraphics[width=0.225\columnwidth]{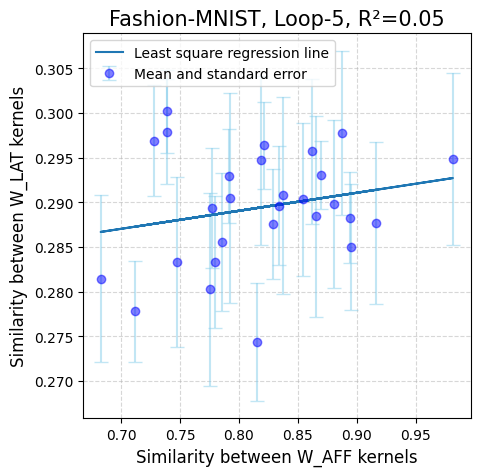}&
~
        \includegraphics[width=0.225\columnwidth]{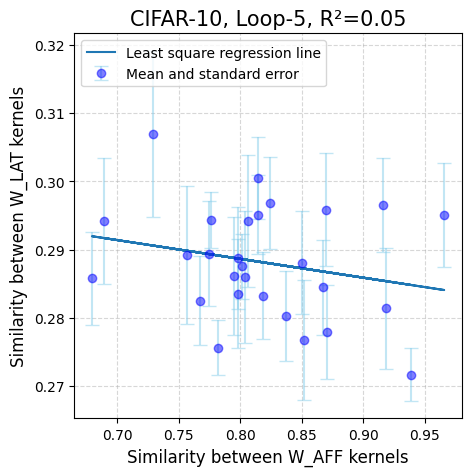}&
~
        \includegraphics[width=0.225\columnwidth]{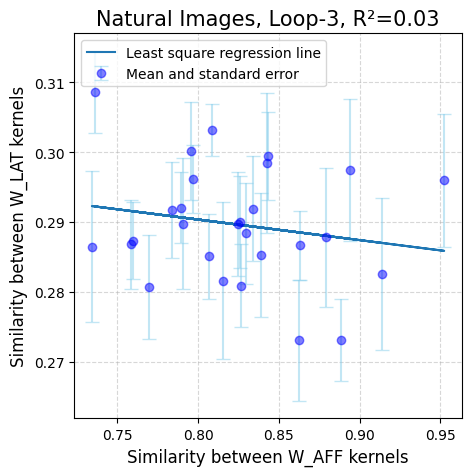}\\
\rotatebox{90}{~~~~~~~~{\normalsize after}} & 
        \includegraphics[width=0.225\columnwidth]{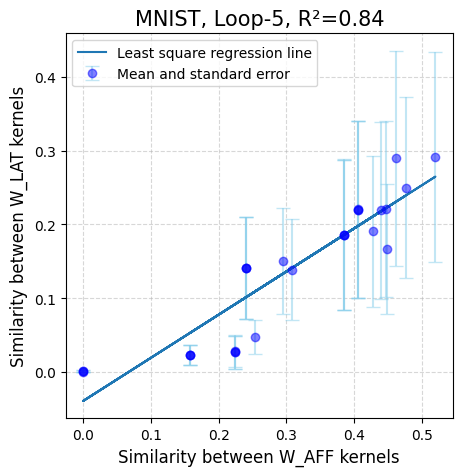} &
~
        \includegraphics[width=0.225\columnwidth]{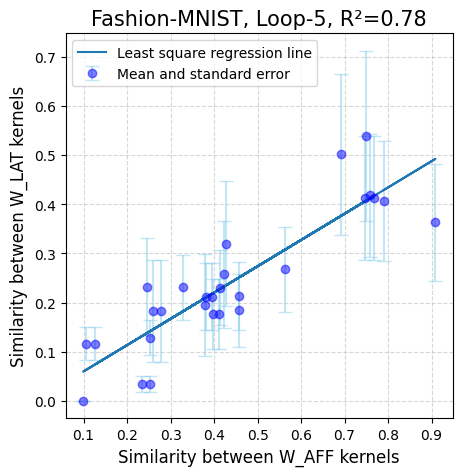} &
~
        \includegraphics[width=0.225\columnwidth]{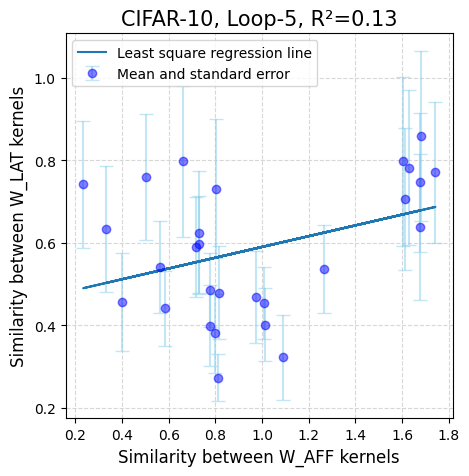} &
~
        \includegraphics[width=0.225\columnwidth]{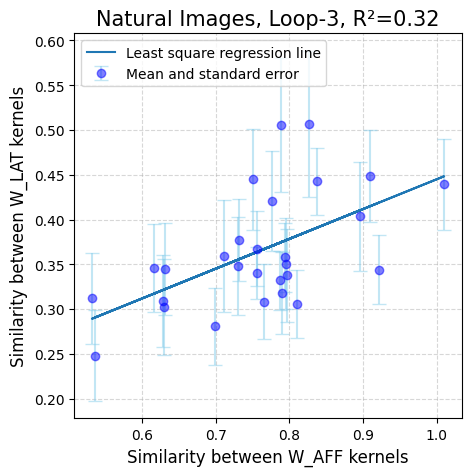} \\
\end{tabular}
\commentout{
    \begin{subfigure}{0.22\columnwidth}
        \centering
        \includegraphics[width=\linewidth]{images/sim_mnist.png}
    \end{subfigure}
    \hspace{0.02\columnwidth} % Adjusts horizontal spacing between columns
    \begin{subfigure}{0.22\columnwidth}
        \centering
        \includegraphics[width=\linewidth]{images/sim_fmnist.png}
    \end{subfigure}
    \hspace{0.02\columnwidth} % Adjusts horizontal spacing between columns
    \begin{subfigure}{0.22\columnwidth}
        \centering
        \includegraphics[width=\linewidth]{images/sim_cifar.png}
    \end{subfigure}
    \hspace{0.02\columnwidth} % Adjusts horizontal spacing between columns
    \begin{subfigure}{0.22\columnwidth}
        \centering
        \includegraphics[width=\linewidth]{images/sim_natural.png}
    \end{subfigure}
}
    \caption{Similarity between convolution kernels of channels in $W_{\textit{AFF}}$ and the corresponding slices in the $W_{\textit{LAT}}$ tensors was assessed for each dataset before (top row) and after training (bottom row). Each data point shows the mean and standard error of the similarity from multiple $W_{\textit{LAT}}$ channels. $R^2$ value was derived using least square regression.}
    \label{fig: exp1 similarity}
\end{figure}

Fig.\ \ref{fig: weight sim des} shows how the relationship between afferent ($W_{\textit{AFF}}$) and lateral weights ($W_{\textit{LAT}}$) can be analyzed, whether similar correspondence exists as in Bosking et al.'s work \cite{bosking1997orientation} (Fig.\ \ref{f:bosking}). The basic idea is that feature maps in different channels in the middle with similar afferent weight properties (i.e., similar convolution kernels in $W_{\textit{AFF}}$) should have similar outgoing lateral connection weight properties (i.e., similar convolution kernels in $W_{\textit{LAT}}$). 
For this, we check the relationship between (1) the similarity in the pair of kernels in $W_{\textit{AFF}}$ and (2) the similarity in the pair of corresponding slices in the $W_{\textit{LAT}}$ tensor (indexed $p$ and $q$). We compute the Euclidean distance between the convolutional kernels for channels \textit{p} and \textit{q} in $W_{\textit{AFF}}$, and do the same for slices \textit{p} and \textit{q} in the $W_{\textit{LAT}}$ tensor of each lateral activation channel. This is computed over all pairs of \textit{p} and \textit{q} ($p \ne q$). (See SM \ref{s:sim} for details.)
The results are shown in Fig.\ \ref{fig: exp1 similarity}. We can see a clear linear relationship, suggesting that feature maps that are based on similar orientation preferences have similar outgoing lateral connection patterns. This is in line with the observed lateral connection properties in Fig.\ \ref{f:bosking}.
\begin{table*}[!htbp]
    \centering
    \begin{tabular*}{\textwidth}{@{\extracolsep{\fill}}lcccc@{}}
        \toprule
        Option / Dataset & MNIST & Fashion-MNIST & CIFAR-10 & Natural Images\\
        \toprule
        (Baseline) No Penalty      & 97.92 $\pm$ 0.20 & 88.22 $\pm$ 0.42 & 50.80 $\pm$ 4.41 & 79.58 $\pm$ 1.08 \\
        \midrule
        (Ours) ABS              & \text{\: }97.86 $\pm$ 0.17* & 88.44 $\pm$ 0.26 & \text{\: }50.42 $\pm$ 1.94* & 81.46 $\pm$ 1.68 \\
        (Ours) SAT-ABS          & \textbf{98.14 $\pm$ 0.05} & 89.48 $\pm$ 0.30 & \textbf{53.52 $\pm$ 1.14} & \textbf{81.48 $\pm$ 0.24} \\
        (Ours) EXP-SAT          & 98.20 $\pm$ 0.12 & \textbf{89.58 $\pm$ 0.24} & 52.76 $\pm$ 0.78 & 80.76 $\pm$ 1.29 \\
        (Ours) SAT-EXP          & 98.04 $\pm$ 0.11 & 89.00 $\pm$ 0.21 & 50.98 $\pm$ 1.03 & 81.04 $\pm$ 0.93 \\
        \bottomrule
    \end{tabular*}
    \caption{Comparison of test accuracies between the baseline and different weight separation loss/penalty terms in LCEI-CNN. The mean test accuracy and standard deviation are computed for each experiment across five runs. The best accuracy is marked in bold.}
    \label{table: exp2 dual pathway sp vs dp}
\end{table*}
\subsection{Model 2: Excitatory and Inhibitory Separation in Laterally Connected CNN (LCEI-CNN)}
\paragraph{Performance:} 
For the second experiment, we evaluated the performance of the Laterally Connected LCEI-CNN. As a baseline model for comparison, we prepared LCEI-CNN without weight separation penalty ($\lambda_2=0$). An analysis of the results shown in Table \ref{table: exp2 dual pathway sp vs dp} reveals that the LCEI-CNN with excitatory/inhibitory separation outperforms the baseline regardless of the choice of the weight separation loss function, except for two cases: ABS MNIST and ABS CIFAR-10. These are annotated with an asterisk(*) in Table \ref{table: exp2 dual pathway sp vs dp}. 
\textcolor{black}{We suspect that the poor performance of ABS is due to the lack of a saturation process, unlike other penalization methods.
Overall,} this indicates that having separate excitatory and inhibitory neuronal populations may have a performance advantage. 

\begin{figure}[!hbp]
    \centering
    \includegraphics[width=0.95\columnwidth]{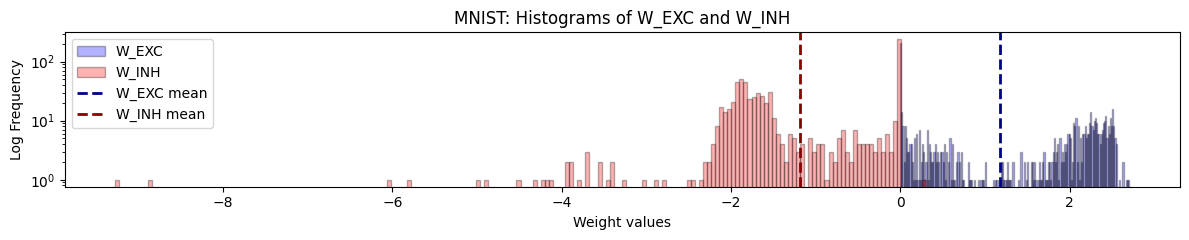}
    \includegraphics[width=0.95\columnwidth]{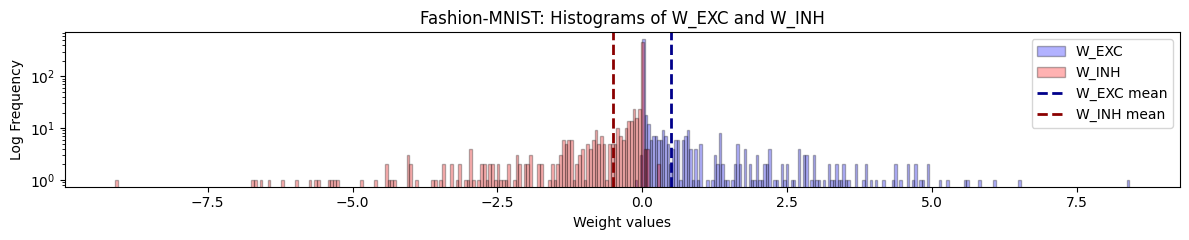}
    \includegraphics[width=0.95\columnwidth]{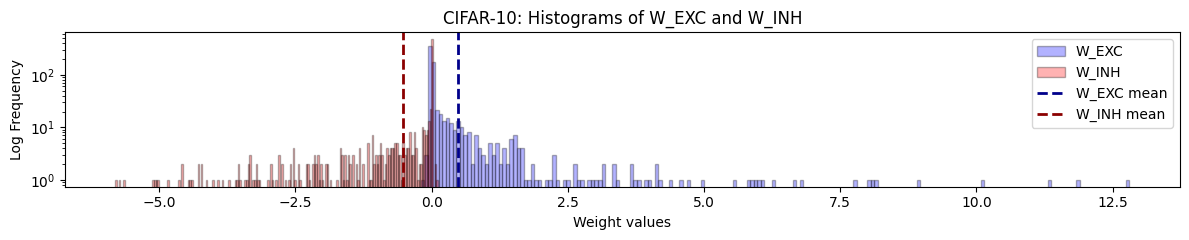}
    \includegraphics[width=0.95\columnwidth]{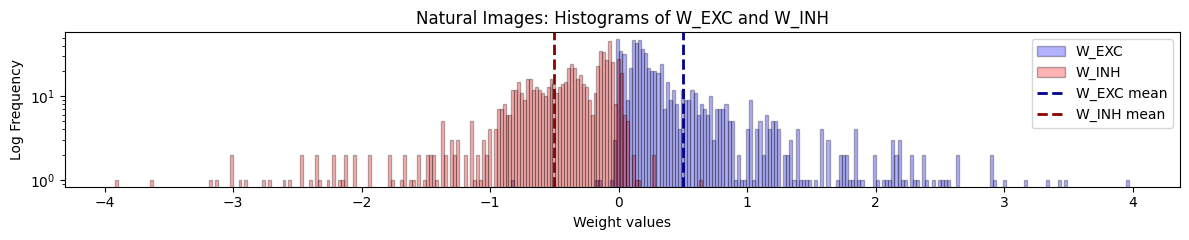}
    \caption{Top: The weight distributions of $W_{\textit{EXC}}$ and $W_{\textit{INH}}$ of the LCEI-CNN trained on the MNIST (SAT-ABS), Fashion-MNIST (EXP-SAT), CIFAR-10 (SAT-ABS), and Natural Images (SAT-ABS).}%Bottom: The feature maps of $O_{\textit{AFF}}$, $O_{\textit{INH}}$, $O_{\textit{EXC}}$, and $O_{\textit{LAT}}$ from EI-CNN: SAT-ABS trained on Natural Images. Each feature map was summed into a single feature map by summing the maps across the channel axis. The same image is used from Fig.\ \ref{fig: exp1_loop}.}
    \label{fig: fmnist weight}
\end{figure}

\begin{figure}[!htbp]
    \centering
    \includegraphics[width=0.85\columnwidth]{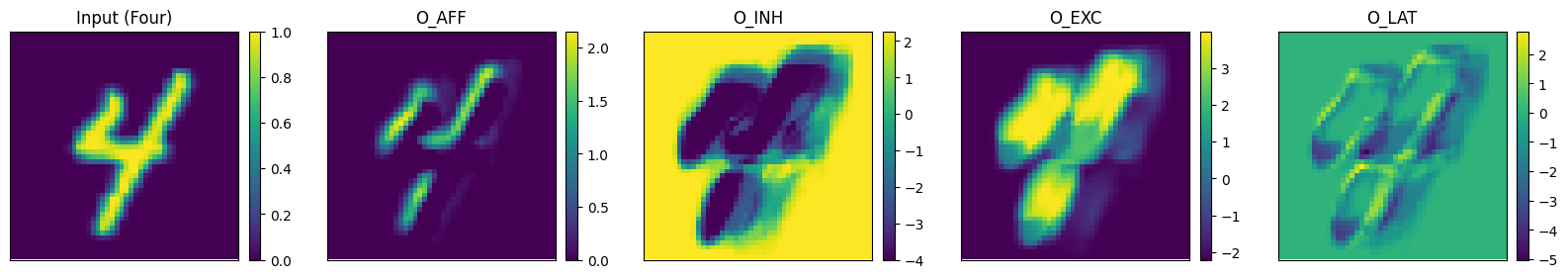}
    \includegraphics[width=0.85\columnwidth]{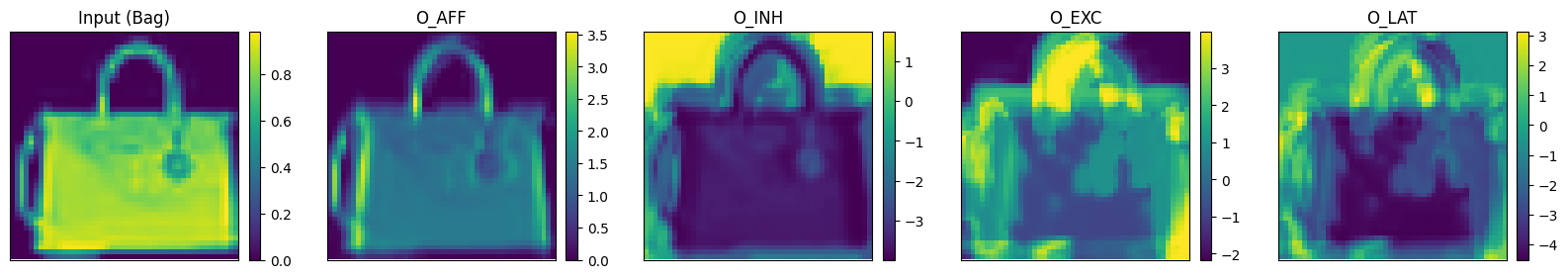}
    \includegraphics[width=0.85\columnwidth]{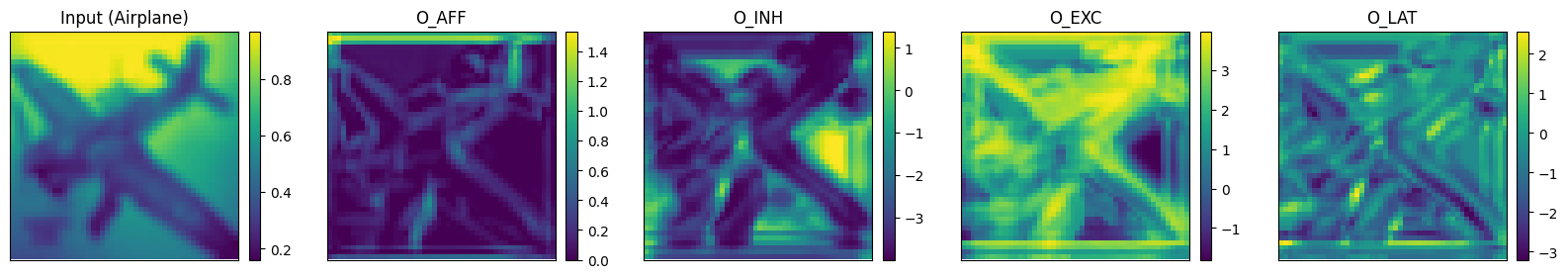}
    \includegraphics[width=0.85\columnwidth]{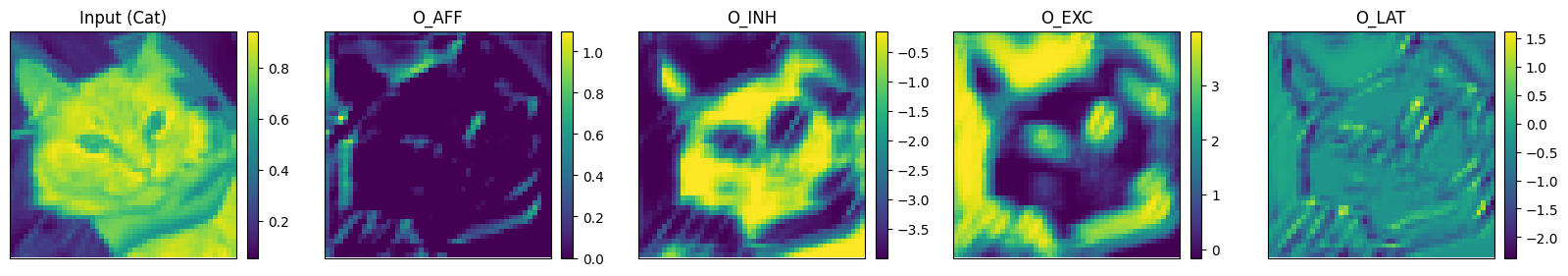}
    \caption{The feature maps of LCEI-CNN. First column = input. Second to fifth column are feature maps of $O_{\textit{AFF}}$, $O_{\textit{INH}}$, $O_{\textit{EXC}}$, and $O_{\textit{LAT}}$. Top-to-Bottom: MNIST (SAT-ABS), FMNIST (EXP-SAT), CIFAR-10 (SAT-ABS), and Natural Image (SAT-ABS). }
    \label{fig: fm ei}
\end{figure}

% yschoe: This is a bit trivial. 
%Comparing the without weight separation loss (No penalty) option and the best option, it can be seen that the test accuracy differences in CIFAR-10 and Natural Images are greater than those for MNIST and Fashion-MNIST. We suspect this is because CIFAR-10 and Natural Images are more challenging datasets.
% yschoe: Kind of redundant.
%By observing the result in the table, it is evident that the application of a weight separation loss that encourages excitatory-inhibitory weight separation contributes to performance enhancement. 
%yschoe: I think it'd be hard to argue this.
%We believe that implementing a weight separation penalty not only emulates the biological properties of excitatory and inhibitory lateral connections but also enables the EI-CNN with penalty to explore more optimal local minima compared to the EI-CNN without penalty. 
% yschoe : too speculative
% This effectiveness may arise because the weight separation loss encourages the network to push one group of weights towards positive values and the other towards negative, increasing the opportunity to find new and better local solutions.
% yschoe: this is what's expected, so it is not so surprising.
\paragraph{Analysis:}
To check if the weight separation loss did in fact shift the weight distribution, we plotted the resulting weight distributions (Fig.\ \ref{fig: fmnist weight}). We observed that the penalty term effectively separated one group of weights into positive values ($W_{\textit{EXC}}$) and the other into negative values ($W_{\textit{INH}}$). All weights started near zero at initialization and gradually diverged during training. The pathways for positive and negative weights result in feature maps with mostly positive and mostly negative values (see $O_{\textit{INH}}$ and $O_{\textit{EXC}}$ in Fig.~\ref{fig: fm ei}). 

Another interesting property we can check is the relative proportion of excitatory vs.\ inhibitory neurons in the cortex, which is known to be around 8:2 or 7:3 \cite{markram2004interneurons,sahara2012fraction}. 
To check if this is the case, we computed the excitatory-to-inhibitory ratio in our trained LCEI-CNN. Table \ref{table: exp2 ratio} shows the ratio of strictly positive ($w > \theta$) to strictly negative ($w < -\theta$) weight values in $W_{\textit{EXC}}$ and $W_{\textit{INH}}$, respectively (with different threshold $\theta$). However, as it can be seen, the proportions are closer to 0.5 : 0.5. This requires further investigation, since in our case, we are counting connections, and the number of neurons and the number of connections may not exactly match.

\commentout{
\begin{figure}[!htbp]
    \centering
    \begin{subfigure}{\textwidth}
        \centering % Centers the subfigure content
    \end{subfigure}
    \novspace{0.01cm}
    \begin{subfigure}{\textwidth}
        \centering % Centers the subfigure content
    \end{subfigure}
    \novspace{0.01cm}
    \begin{subfigure}{\textwidth}
        \centering % Centers the subfigure content
    \end{subfigure}
    \caption{The distribution of $W_{\textit{EXC}}$ and $W_{\textit{INH}}$ of the LCEI-CNN: SAT-ABS, LCEI-CNN: EXP-SAT, and LCEI-CNN: SAT-ABS trained on the MNIST, Fashion-MNIST, and CIFAR-10, respectively.}
\end{figure}
}

\begin{table}[!htbp]
  \begin{center}
  \begin{tabular*}{\columnwidth}{@{}l@{}c@{}c@{}c@{}c@{}}
    \toprule
    Thr/Dat & MNIST & F-MNIST & CIFAR-10 & Nat.\ Img.\\
    \toprule
    $\theta=0$ &~0.505:0.495~ &~0.499:0.501~ &~0.524:0.476~ &~0.505:0.495 \\
    $\theta=1$ &~0.469:0.531~ &~0.500:0.500~ &~0.424:0.576~ &~0.609:0.391 \\
    $\theta=2$ &~0.760:0.240~ &~0.527:0.473~ &~0.388:0.612~ &~0.578:0.422 \\
    \bottomrule
  \end{tabular*}
  \end{center}
    \caption{Proportions of strictly excitatory and strictly inhibitory weight values ($\theta$ = threshold: $w > \theta$ or $w < -\theta$).}
  \label{table: exp2 ratio}
\end{table}

% yschoe: again, these are lateral connections, not afferent connections, so Gabor patterns are not expected.
% We identified some patterns in $W_{\textit{EXC}}$ and $W_{\textit{INH}}$ exhibiting patterns of edge detectors (see Fig.\ \ref{fig: penalty gabor pattern}), albeit not as pronounced as in the first experiment. Note that we used the same dataset from the first experiment for better comparison. Comparing Fig.\ \ref{fig: exp1_weight kernel} and \ref{fig: penalty gabor pattern}, the weight kernels from the first experiment exhibit a closer resemblance to Gabor filters than those from the second experiment. 
\section{Discussion}

The main contribution of this paper is in the introduction of the concept of lateral connections into CNN design, and the analysis of response and connection properties in the context of the biological counterpart. Similar approaches exist such as RCNN \cite{liang2015recurrent} but they focused more on performance and the recursive aspect. Through our analysis, we found that sparseness and power-law-like response characteristics in biology and biologically accurate models can be achieved with only lateral connections and standard gradient-based learning in CNN. This is interesting compared to related works since we did not include any explicit activity sparsification terms as in \cite{bizopoulos2020sparsely}, and did not use Hebbian learning as in \cite{miikkulainen2006computational}. In terms of lateral connection characteristics, as mentioned already, our analysis and its interpretation are limited due to the weight-sharing in the convolution operation inherent in CNN. In future work, we can alleviate this by removing weight sharing in CNN, as proposed by \cite{bartunov2018assessing}. We also need further analysis of the lateral connection weights for our LCEI-CNN model. Analyzing the spectral properties (e.g., eigenvalue distribution of the weights) may give us important insights. For example, \cite{li2023learning} showed that spectral properties of the weight matrices matter more than strict constraints on the sign of the weights. 
Also, we plan to combine model 1 and model 2 into a singe model, and expand our approach to spiking CNN. 

\section{Conclusion} The main novelty of this paper is the use of lateral connections in CNN, inspired by the biological visual system, as a new architectural component in convolutional neural networks.  Unlike afferent connections and long-range projections, the equivalence of which is already present in CNN, lateral connections establishing local connections within a visual cortical area have no counterpart in the existing CNN models. In CNN,  lateral connections can be implemented as connections within a feature map and across feature maps in the same convolutional layer. We tested two new CNN models that incorporate these lateral connections, and tested two main properties: (1) recurrent activation through lateral connections, and (2) separation of excitatory and inhibitory lateral connections. We observed that in both cases, classification accuracy increased compared to the baseline. Furthermore, we found several emergent structural and functional properties in our laterally connected CNN that parallel known observations in the neuroscience literature. These include the sparsification through recurrent activation, and lateral connection properties aligning with the afferent stimulus specificity. 
We expect our work to help understand the computational role of lateral connections in the visual cortex, and also build more powerful biologically inspired CNN architectures.  

\newpage
\bibliographystyle{apalike}
\bibliography{references}   % Your .bib file (without .bib extension)
\clearpage
\appendix

\onecolumn

\section{Supplementary Materials (SM)} \label{section: sm}
\subsection{Network Architectures}
\label{s:archi}
\begin{table}[!htbp]
    \caption{The architecture of LC-CNN. This configuration applies to all LC-CNN structures with different numbers of loops. FM denotes the feature map, and the stride is 1. The input is a 48$\times$48 grayscale image. The input is followed by $O_{\textit{AFF}}$ and $O_{\textit{LAT}}$.}
  \begin{center}
  \begin{tabular*}{\columnwidth}{@{\extracolsep{\fill}}c|cccc@{}}
    \toprule
    Layer name & Output FM size & Input $\rightarrow$ Output channel depth & Kernel size & Activated by\\
    \midrule
    $O_{\textit{AFF}}$   & 48$\times$48 & 1 $\rightarrow$ 8 & 7$\times$7 & ReLU\\
    $O_{\textit{LAT}}$   & 48$\times$48 & 8 $\rightarrow$ 8 & 7$\times$7 & ReLU\\
    Max-pooling & 24$\times$24 & 8 $\rightarrow$ 8 & 2$\times$2 & -\\
    \bottomrule
  \end{tabular*}
  \end{center}
  \label{table: exp1 archi design}
\end{table}
% Add code snippet here, if you think it is appropriate. \marginnote{Add code snippet if appropriate.}
% \begin{verbatim}
% # Code snippet : TensorFLow, LeNET -- replace this
% model = Sequential()
% model.add(Conv2D(25, (7, 7), activation='relu', input_shape=(28, 28, 1)))
% model.add(MaxPooling2D((2, 2)))
% model.add(Conv2D(25, (7, 7), activation='relu'))
% model.add(MaxPooling2D((2, 2)))
% model.add(Flatten())
% model.add(Dense(256, activation='relu'))
% model.add(Dense(10, activation='softmax'))
% \end{verbatim}
\begin{table}[!htbp]
  \caption{The architecture of LCEI-CNN. This configuration applies to all LCEI-CNN structures with different weight separation loss (penalty) options. FM denotes the feature map, and the stride is 1. The input is a 48$\times$48 grayscale image. The input is followed by $O_{\textit{AFF}}$, $O_{\textit{EXC}}$ and $O_{\textit{INH}}$ together, and $O_{\textit{LAT}}$.}
  \begin{center}
  \begin{tabular*}{\columnwidth}{@{\extracolsep{\fill}}c|cccc@{}}
    \toprule
    Layer name & Output FM size & Input $\rightarrow$ Output channel depth & Kernel size & Activated by\\
    \midrule
    $O_{\textit{AFF}}$   & 48$\times$48 & 1 $\rightarrow$ 4 & 7$\times$7 & ReLU\\
    $O_{\textit{EXC}}$   & 48$\times$48 & 4 $\rightarrow$ 4 & 7$\times$7 & Tanh\\
    $O_{\textit{INH}}$   & 48$\times$48 & 4 $\rightarrow$ 4 & 7$\times$7 & Tanh\\
    $O_{\textit{LAT}}$   & 48$\times$48 & 4 $\rightarrow$ 4 & - & $\oplus$ \\
    Max-pooling & 24$\times$24 & 4 $\rightarrow$ 4 & 2$\times$2 & -\\
    \bottomrule
  \end{tabular*}
  \end{center}
  \label{table: exp2 archi design}
\end{table}
% Add code snippet here, if you think it is appropriate.
% \marginnote{Add code snippet if appropriate.}
% \begin{verbatim}
% # Code snippet : TensorFLow, LeNET -- replace this
% model = Sequential()
% model.add(Conv2D(25, (7, 7), activation='relu', input_shape=(28, 28, 1)))
% model.add(MaxPooling2D((2, 2)))
% model.add(Conv2D(25, (7, 7), activation='relu'))
% model.add(MaxPooling2D((2, 2)))
% model.add(Flatten())
% model.add(Dense(256, activation='relu'))
% model.add(Dense(10, activation='softmax'))
% \end{verbatim}
\subsection{Computing resources}
\label{s:compute}
% \marginnote{Add computing resources: CPU, GPU, typical training time, etc.}
We utilized an Intel i9-13900HX CPU and an RTX 4070 Laptop GPU. For each experiment, the training time for LC-CNN typically ranges from 1 to 2 hours, while LCEI-CNN requires approximately 3 to 4 hours, with 4 to 5 instances of the code running in parallel. However, these durations are influenced by the learning rate and scheduler. Please see the code files for more details.

\subsection{Solving $c\frac{1}{x^{a}} = \frac{1}{\sigma\sqrt{2\pi}} e^{-\frac{x^2}{2\sigma^2}}$ (Sketch)}
\label{s:solve}
Start with
\[ c \frac{1}{x^{a}} = \frac{1}{\sigma\sqrt{2\pi}} e^{-\frac{x^2}{2\sigma^2}}. \]
Rearrange to get
\[ c \sigma\sqrt{2\pi} = x^{a} e^{-x^2/(2\sigma^2)}. \]
Let
\[ u = \frac{x^2}{2\sigma^2}, \]
and rearrange to get
\[ c \sigma\sqrt{2\pi} = (2\sigma^2)^{a/2} u^{a/2} e^{-u}. \]
Isolate the $u^{a/2}$ term to the left and raise both sides to the power of $2/a$ to get
\[ u = (c \sigma\sqrt{2\pi}(2\sigma^2)^{-a/2})^{2/a}e^{2u/a},\]
then multiply both sides with $e^{-2u/a}$ to get
\[ u e^{-2u/a} =  (c \sigma\sqrt{2\pi}(2\sigma^2)^{-a/2})^{2/a}.\]
Now we have a rough form where the Lambert $W$ function can be applied, but we need one more step.
Let
\[ y = - \frac{2u}{a}, \]
then
\[ u = -\frac{a}{2}y,\]
and, after a few simple steps we get a form suitable for the application of the Lambert $W$ function:
\[ y e^y = - \frac{2}{a}(c \sigma\sqrt{2\pi}(2\sigma^2)^{-a/2})^{2/a}.\]
Simplifying the constants and applying the Lambert $W$ function gives
\[ y = W_k\left( -\frac{(c \sigma\sqrt{2\pi})^{2/a}}{a\sigma} \right),\]
where $k$ identifies the branch of $W$ (0=principal branch, -1=lower real branch).
Subsituting back $y$ and $u$ and rearranging, we get the final result:
\[x =    \pm \sqrt{ - a\sigma^2 W\left( -\frac{(c \sigma\sqrt{2\pi})^{2/a}}{a\sigma^2} \right) }. \]

\subsection{Similarity measure for convolution kernels} \label{exp1 weight similarity}
\label{s:sim}
The similarity value between each possible pair of weight kernels is calculated by the Euclidean distance. Assuming there are $k$ weight kernels for $W_{\textit{AFF}}$, this results in $\comb{\textit{k}}{2}$ different similarity measurements as the input image is grayscale. The similarity value between the $p$-th and $q$-th ($p \neq q$) weight kernels can be measured using Eq/\ \ref{similarity} where $n$ denotes the square of kernel size. 
\begin{align} \label{similarity}
    sim(p, q) = \sqrt{\sum_{i=1}^n (p_i - q_i)^2}
\end{align}
Assuming the input channel has a depth of $k$, and the output channel has a depth of $l$ for $W_{\textit{LAT}}$, the similarity value for the $p$-th and $q$-th weight kernels for each $l$-th weight tensor can also be measured by the above equation (See Eq.\ \ref{similarity}). Note that $k=l$ must hold, as the input and output channel depths for $W_{\textit{LAT}}$ need to be the same. Due to the presence of $l$ different output tensors for $W_{\textit{LAT}}$, we cannot directly compare $W_{\textit{AFF}}$ and $W_{\textit{LAT}}$. We can calculate the mean and standard error (SE) of $l$ different similarity values for each pair of weight kernels (See Eq.\ \ref{similarity2} and \ref{similarity3}).
\begin{align} \label{similarity2}
    \mathbb{E}[sim(p, q)]  &= \frac{1}{l} \sum_{j=1}^l sim_j(p, q) \\
    \label{similarity3}
    SE[sim(p, q)] &= \sqrt{\frac{\mathbb{V}[sim(p, q)]}{l}}
\end{align}
In this manner, for every pair of $p$-th and $q$-th weight kernels in $W_{\textit{AFF}}$ and $W_{\textit{LAT}}$, we can plot a scattered similarity graph using $W_{\textit{AFF}}$'s $sim(p, q)$ and $W_{\textit{LAT}}$'s $\mathbb{E}[sim(p, q)]$. The error boundary is represented by $SE[sim(p, q)]$ (see Fig.\ \ref{fig: exp1 similarity} or SM Fig.\ \ref{fig: exp1 similarity before train}). 

\newpage
\subsection{Supplementary results (Convolution Kernels)} \label{SM additional figures}

\begin{figure}[!htbp]
    \begin{subfigure}{\textwidth}
        \centering % Centers the subfigure content
        \includegraphics[width=0.45\textwidth]{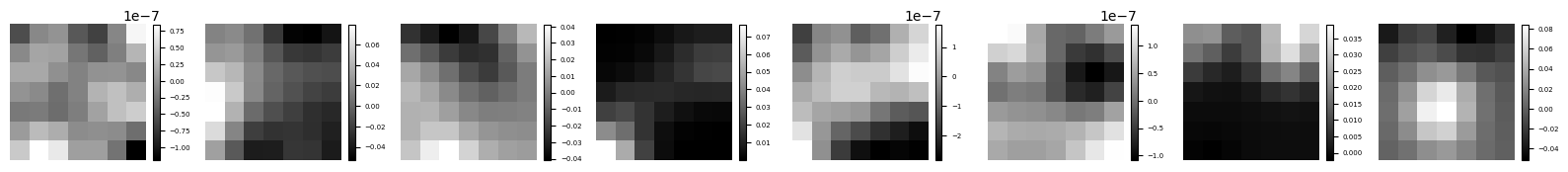}
        \caption{MNIST}
    \end{subfigure}
   % \novspace{0.01\textwidth} % Adds space between the subfigures
    \begin{subfigure}{\textwidth}
        \centering % Centers the subfigure content
        \includegraphics[width=0.45\textwidth]{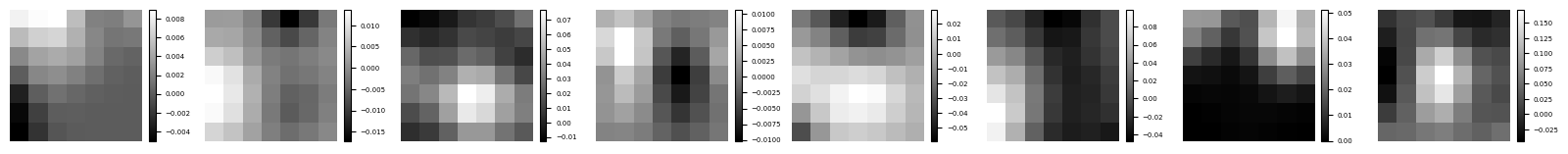}
        \caption{Fashion-MNIST}
    \end{subfigure}
    \novspace{0.01\textwidth} % Adds space between the subfigures
    \begin{subfigure}{\textwidth}
        \centering % Centers the subfigure content
        \includegraphics[width=0.45\textwidth]{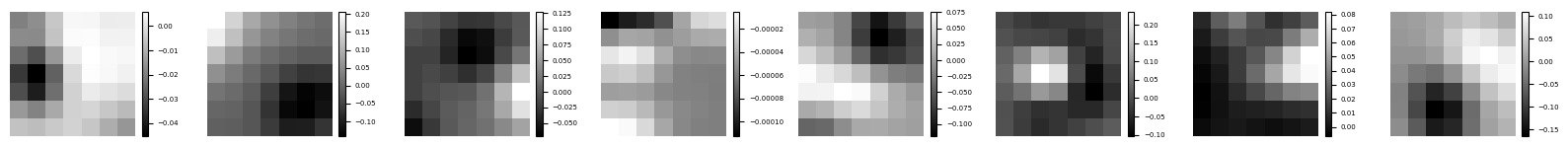}
        \caption{CIFAR-10}
    \end{subfigure}
    \novspace{0.01\textwidth} % Adds space between the subfigures
    \begin{subfigure}{\textwidth}
        \centering % Centers the subfigure content
        \includegraphics[width=0.45\textwidth]{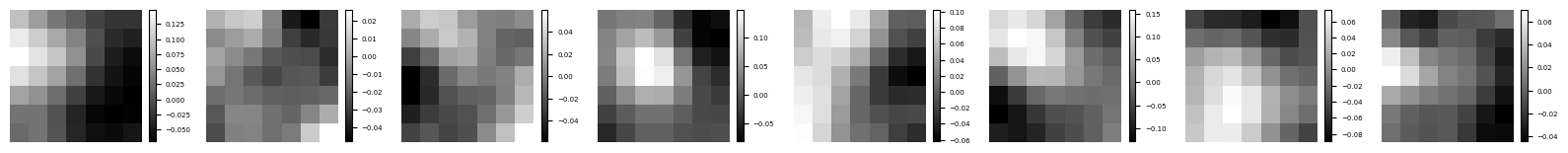}
        \caption{Natural Images}
    \end{subfigure}
    \caption{Afferent weight kernels $W_{\textit{AFF}}$ of Model 1 LC-CNN: MNIST (Loop-5), Fashion-MNIST (Loop-5), CIFAR-10 (Loop-5), and Natural Images (Loop-3) from top to bottom, respectively. Gaussian blur is applied to enhance the visibility. In each subplot, there are 8 channels, from left to right.}
    \label{fig: exp1_weight kernel 3}
\end{figure}

\begin{figure}[!htbp]
    \centering
\begin{tabular}{cc}
        \includegraphics[height=6cm]{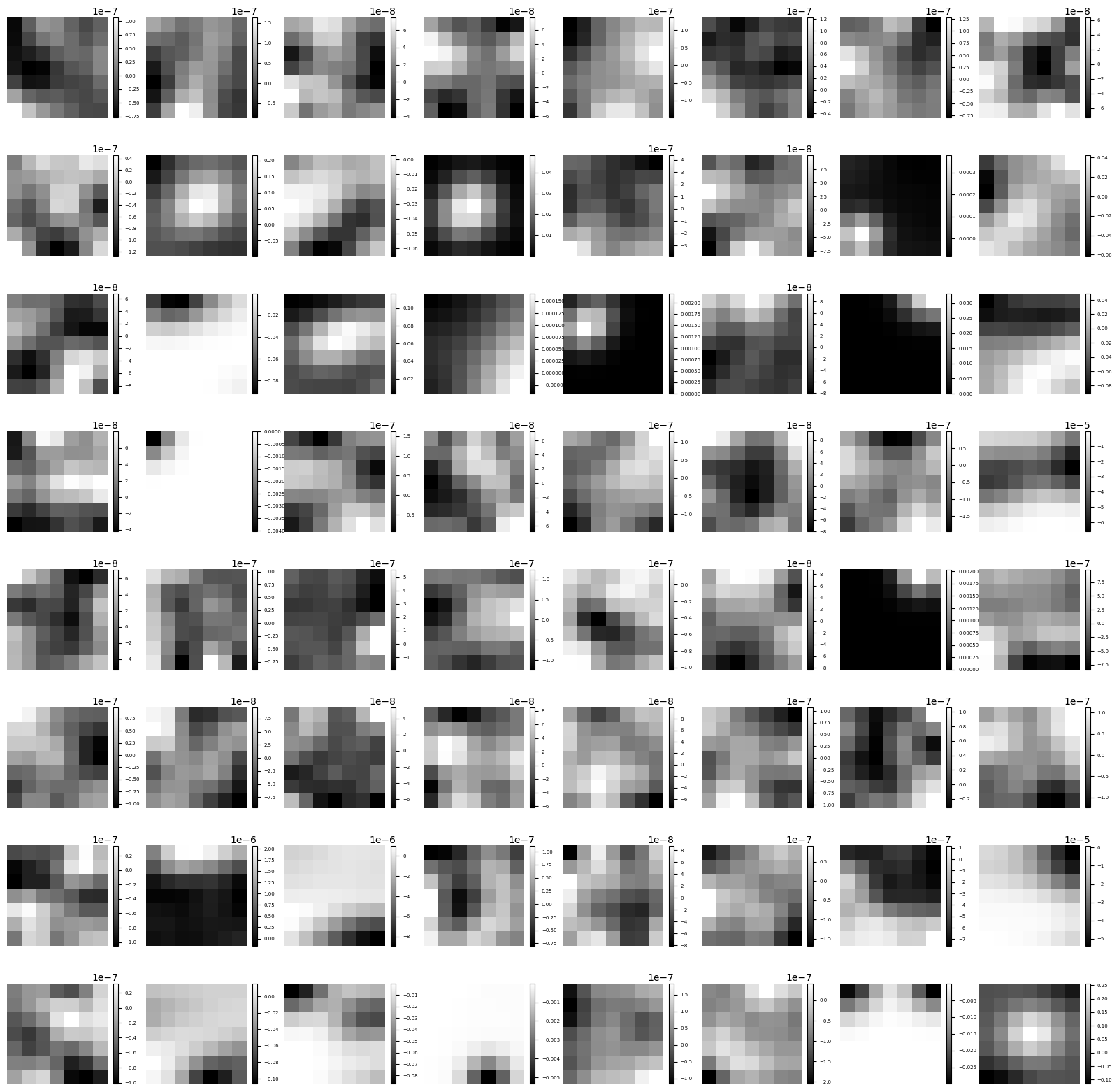}
&
        \includegraphics[height=6cm]{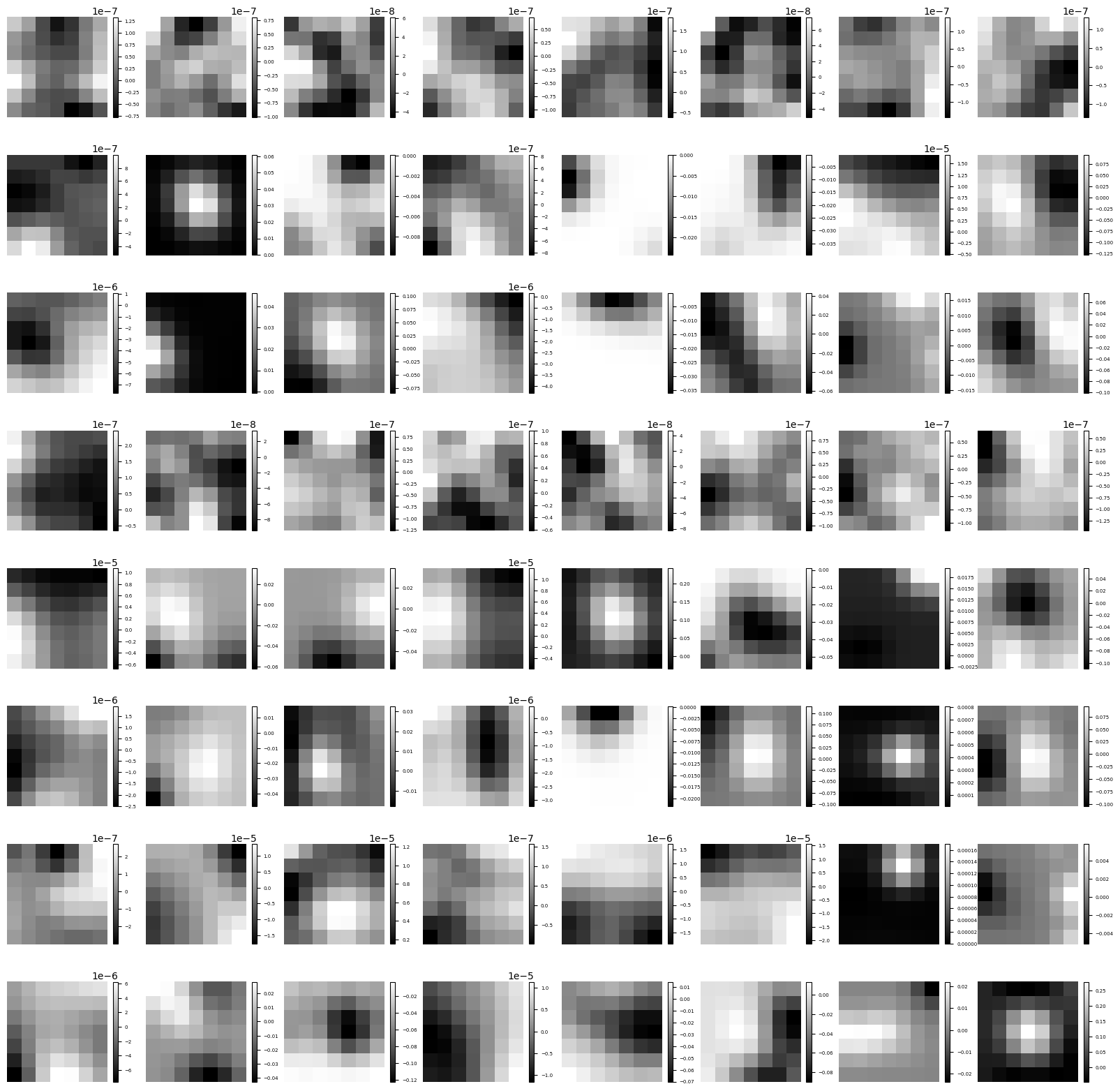}
\\
(a) MNIST & (b) FMNIST \\
        \includegraphics[height=6cm]{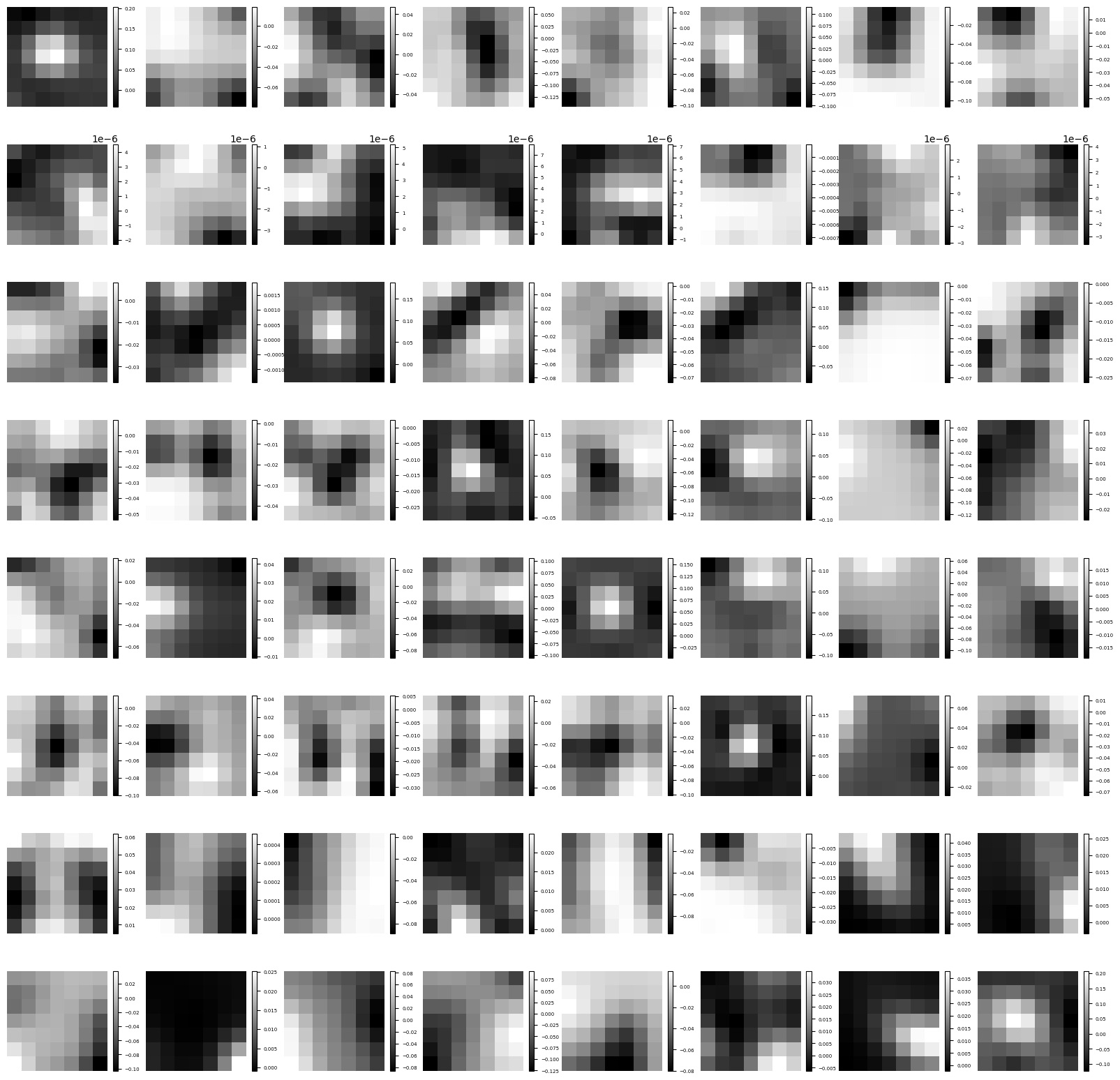} &
        \includegraphics[height=6cm]{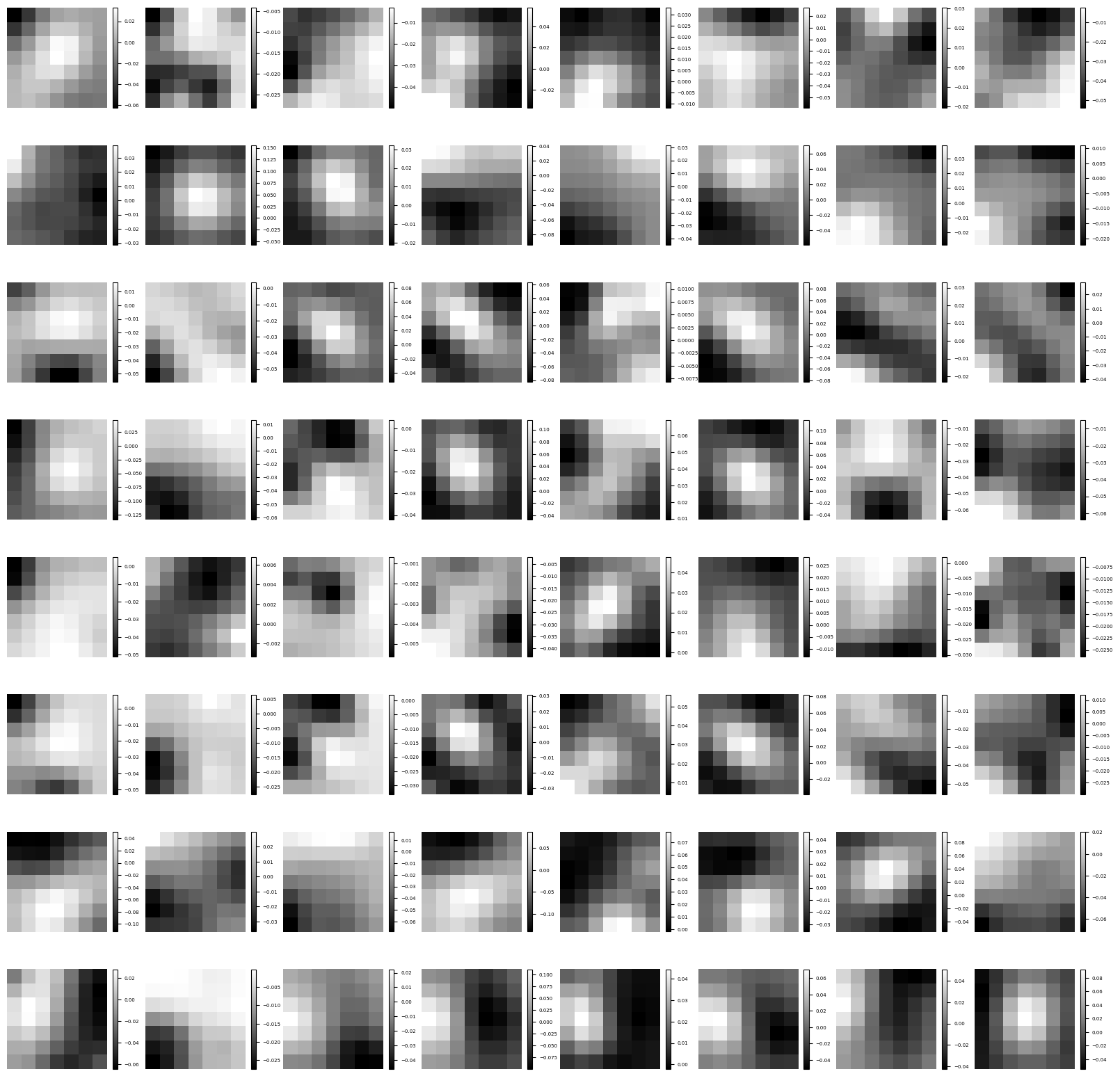}\\
(a) CIFAR-10& (b) Natural\\
\end{tabular}
    \caption{Lateral weight ($W_{LAT}$) kernels of Model 1 LC-CNN. Gaussian blur is applied to enhance the pattern. In each subplot, 8 channels, from top to bottom.}
    \label{fig: exp1_weight kernel 2}
\end{figure}

\begin{figure}[!htbp]
\centering
\begin{tabular}{cc}
        \includegraphics[height=3cm]{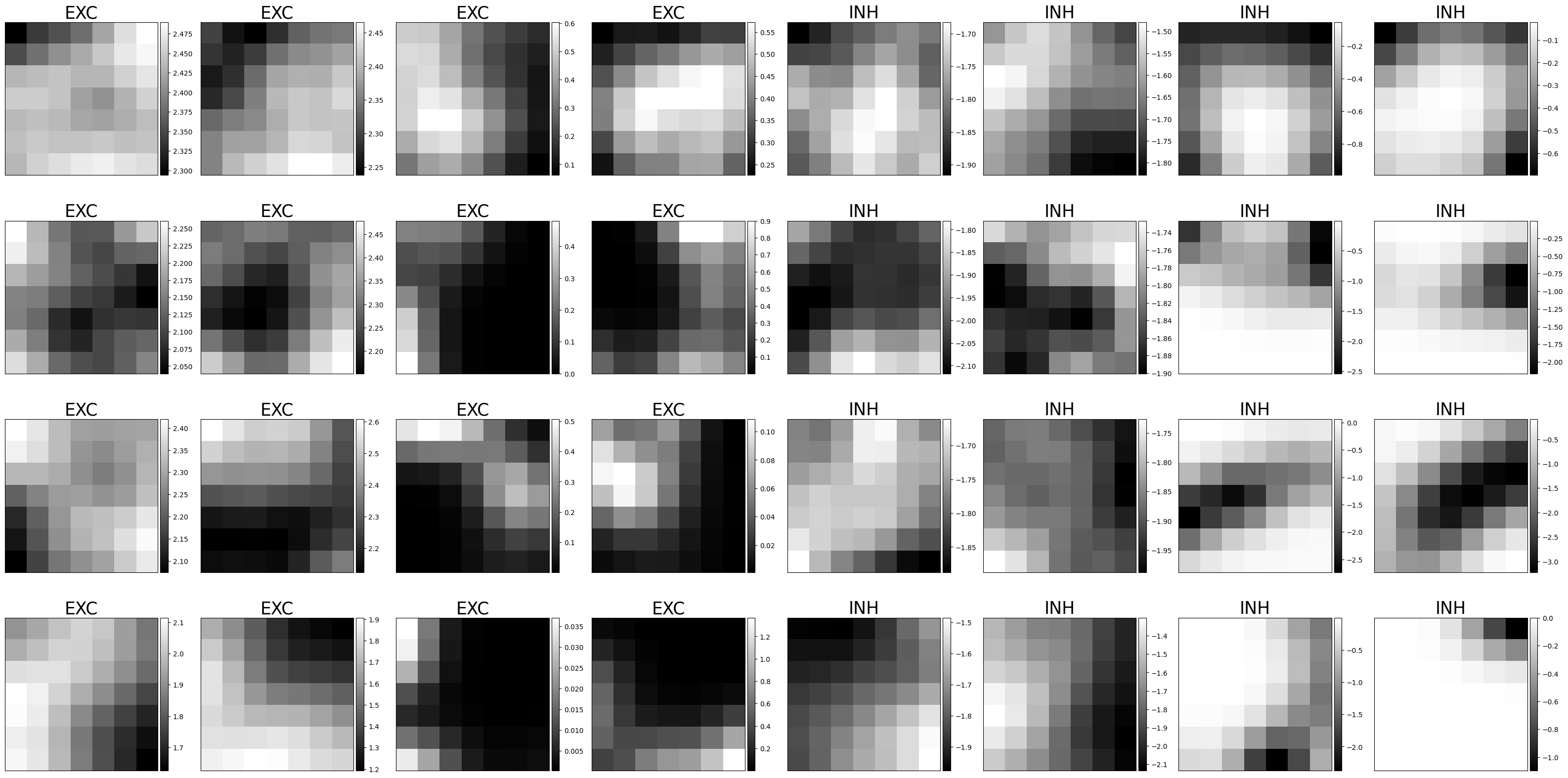} &
        \includegraphics[height=3cm]{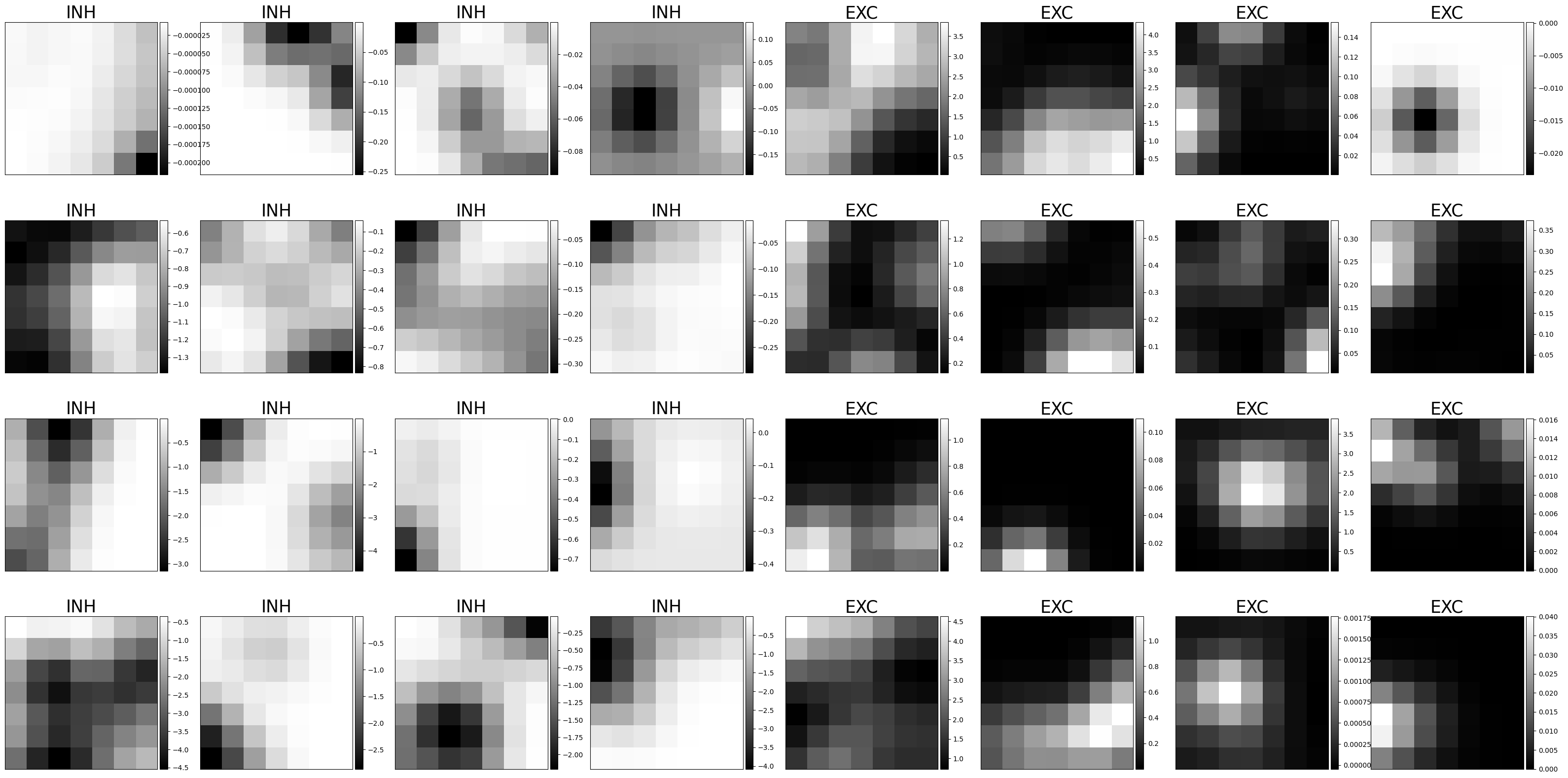} \\
(a) MNIST & (b) FMNIST \\
        \includegraphics[height=3cm]{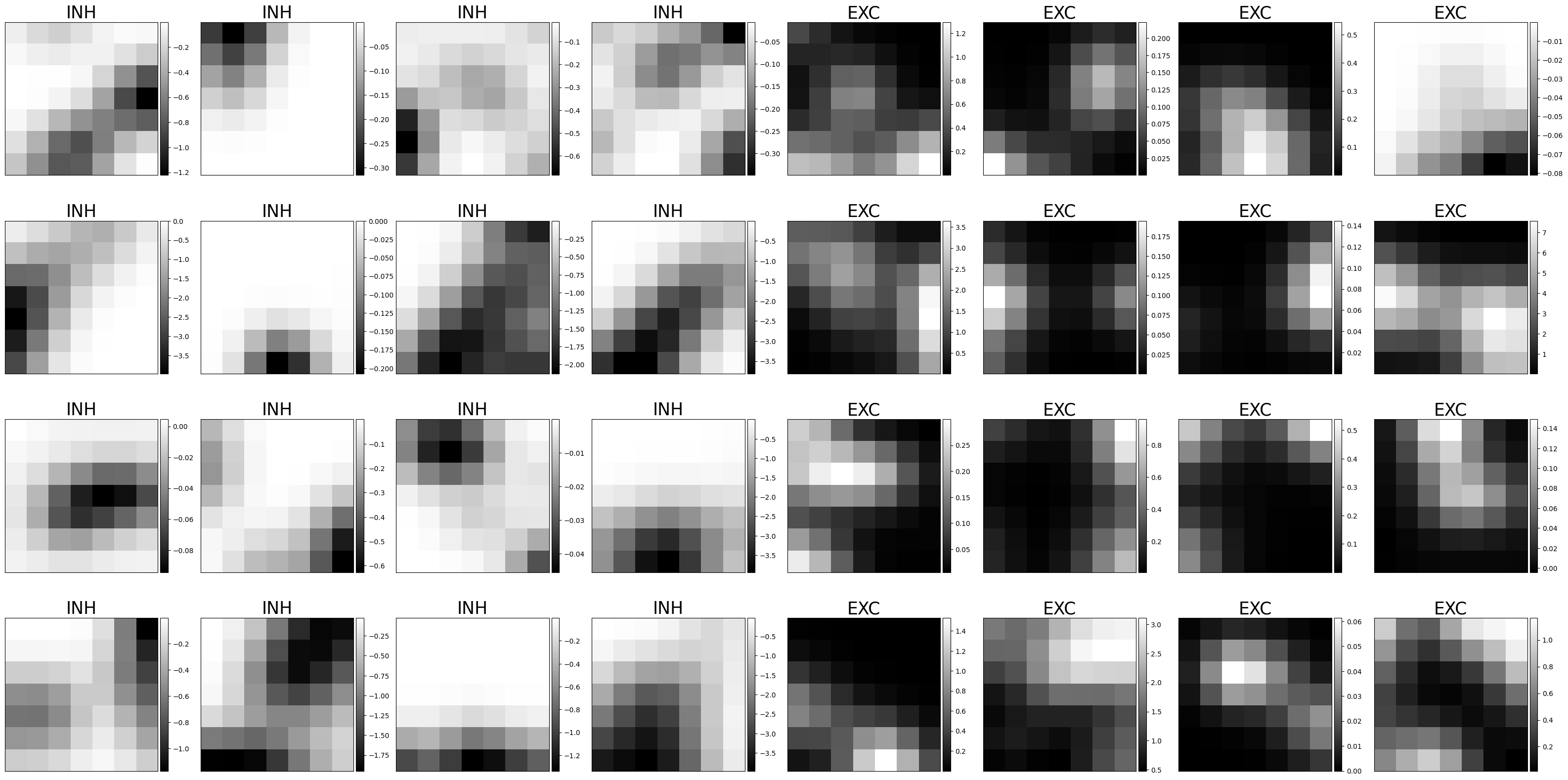} &
        \includegraphics[height=3cm]{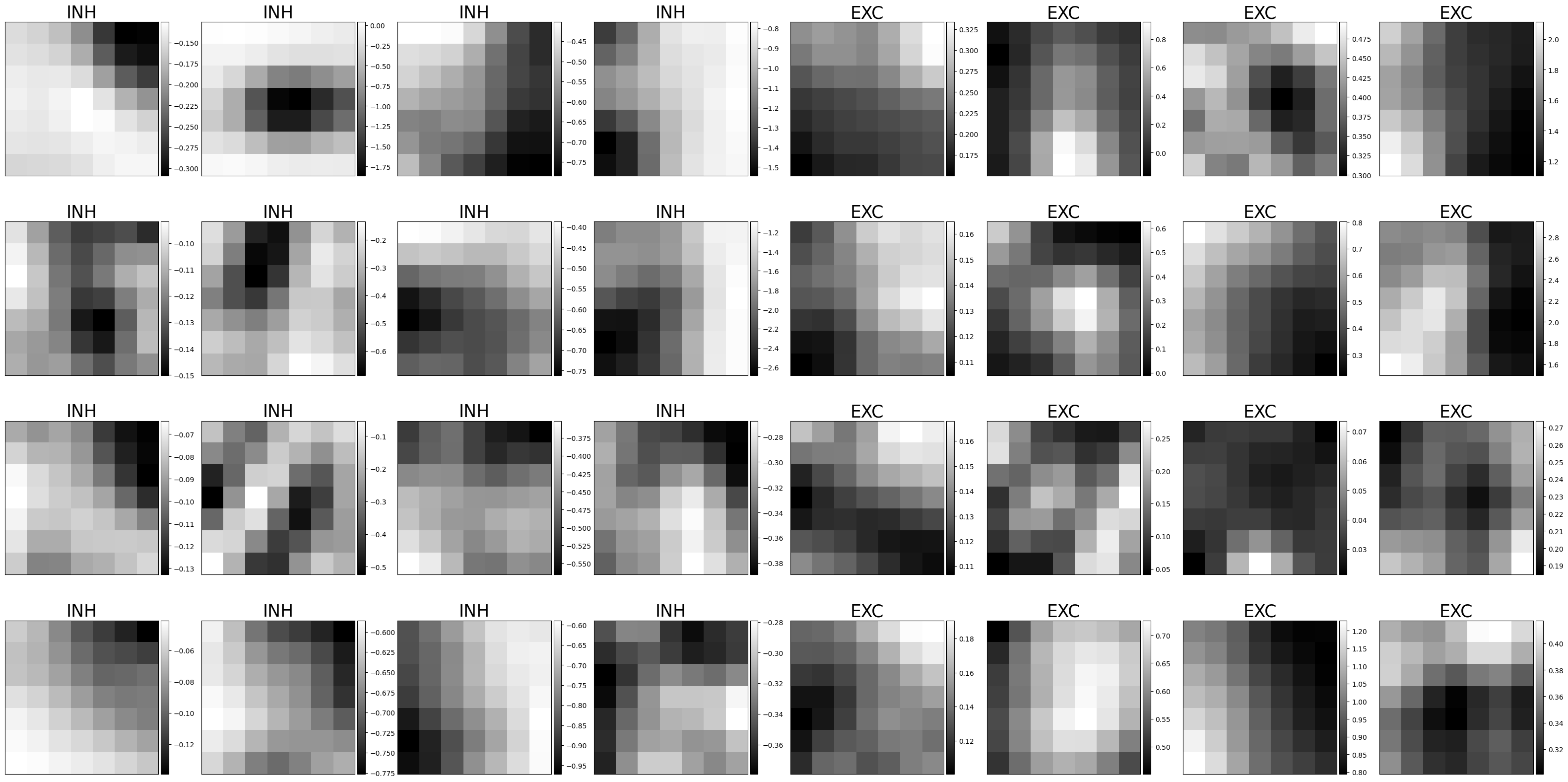}\\
(a) CIFAR-10& (b) Natural\\
\end{tabular}
    \caption{Lateral weight kernels $W_{\textit{EXC}}$ and $W_{\textit{INH}}$ of the model 2 LCEI-CNN: SAT-ABS, EXP-SAT, SAT-ABS, and SAT-ABS trained on the MNIST, Fashion-MNIST, CIFAR-10, and Natural Images, respectively. Gaussian blur is applied to enhance the visibility. In each subplot, there are 4 channels, from top to bottom.}
\end{figure}

\medskip

\clearpage
%%%%%%%%%%%%%%%%%%%%%%%%%%%%%%%%%%%%%%%%%%%%%%%%%%%%%%%%%%%%

\end{document}